\documentclass{article}

\usepackage{arxiv}

\usepackage{appendix}
\appendixtitleon
\appendixtitletocon

\usepackage[utf8]{inputenc} 
\usepackage[T1]{fontenc}    
\usepackage[colorlinks=true,linkcolor=magenta,citecolor=blue,pagebackref]{hyperref}
\usepackage{url}            
\usepackage{booktabs}       
\usepackage{amsfonts}       
\usepackage{amsmath}
\usepackage{nicefrac}       

\usepackage{natbib}
\usepackage{doi}

\usepackage{subfigure, epsfig}

\bibpunct[, ]{(}{)}{,}{a}{}{,}%

\newtheorem{theorem}{Theorem}
\newtheorem{lemma}{Lemma}
\newtheorem{proposition}{Proposition}
\newtheorem{corollary}{Corollary}

\newtheorem{conjecture}{Conjecture}
\newtheorem{hypothesis}{Hypothesis}
\newtheorem{assumption}{Assumption}
\newtheorem{remark}{Remark}
\newtheorem{example}{Example}

\newtheorem{definition}{Definition}

\usepackage{times}
\usepackage{latexsym}
\usepackage{graphicx}
\usepackage{color}
\usepackage{enumitem}

\usepackage{multirow}
\usepackage{threeparttable}

\usepackage{tikz}
\usepackage[labelfont=bf]{caption}
\usepackage{soul}

\usepackage{longtable}
\usepackage{makecell}
\usepackage{multirow}
\usepackage{float}
\usepackage{bbm}

\usepackage{svg}

\usepackage{bibentry}
\usepackage{fancyhdr}
\usepackage{booktabs}
\usepackage{titlesec}
\usepackage{url}

\newcommand{\ind}{{\perp \!\!\! \perp}}
\newcommand{\notind}{{\not\!\perp\!\!\!\perp}}

\newcommand{\ls}[1]
  {\dimen0=\fontdimen6\the\font \lineskip=#1\dimen0
  \advance\lineskip.5\fontdimen5\the\font \advance\lineskip-\dimen0
  \lineskiplimit=.9\lineskip \baselineskip=\lineskip
  \advance\baselineskip\dimen0 \normallineskip\lineskip
  \normallineskiplimit\lineskiplimit \normalbaselineskip\baselineskip
  \ignorespaces }

\iffalse 
\usepackage{amsfonts}
\usepackage{verbatim}
\usepackage{bbm}

\usepackage{graphicx}

\usepackage{authblk}

\usepackage{amsthm}
\newcommand{\field}[1]{\mathbb{#1}}
\DeclareMathOperator{\PR}{\field{P}}             
\DeclareMathOperator{\E}{\field{E}}              
\def\N{\field{N}}                                
\def\R{\field{R}}                                
\def\F{\field{F}}                                

\else

\def\PR{\mathop{\rm I\kern -0.20em P}\nolimits}  
\def\E{\mathop{\rm I\kern -0.20em E}\nolimits}   
\def\N{\mathop{\rm I\kern -0.20em N}\nolimits}   
\def\R{\mathop{\rm I\kern -0.20em R}\nolimits}   
\def\F{\mathop{\rm I\kern -0.20em F}\nolimits}   
\fi

\graphicspath{{./}{Figs/}}

\usepackage{longtable}

\date{\small \textit{\today}}

\title{Evaluation of a Split Flow Model for the Emergency Department}

\author{ \hspace{1mm}Juan Camilo David G\'{o}mez\\
	Department of Industrial and Systems Engineering\\
	University of Wisconsin-Madison\\
	Madison, WI \\
	\texttt{\href{mailto:@wisc.ed}{davidgomez@wisc.edu}} \\
	\And
	\hspace{1mm}Amy Cochran \\
	Department of Population Health Sciences \\ and Department of Mathematics\\
	University of Wisconsin-Madison\\
	Madison, WI \\
	\texttt{\href{mailto:@wisc.ed}{cochran4@wisc.edu}} \\
	\And
	\hspace{1mm}Brian Patterson \\
	BerbeeWalsh Department of Emergency Medicine\\
	University of Wisconsin-Madison\\
	Madison, WI \\
	\texttt{\href{mailto:bapatter@medicine.wisc.edu}{bpatter@medicine.wisc.edu}} \\
	\And
	\hspace{1mm}Gabriel Zayas-Cab\'{a}n \\
	Department of Industrial and Systems Engineering\\
	University of Wisconsin-Madison\\
	Madison, WI \\
	\texttt{\href{mailto:zayascaban@wisc.edu}{zayascaban@wisc.edu}} \\
}

\renewcommand{\shorttitle}{Split Flow}

\hypersetup{
pdftitle={Split flow},
pdfsubject={med},
pdfauthor={David Juan, Cochran Amy, Patterson Brian, Zayas-Caban Gabriel},
pdfkeywords={Split Flow Model; Causal Inference; Emergency Department; Patient Discharge; Electronic Health Records},
}

\begin{document}
\maketitle

\begin{abstract}
	Split flow models, in which a physician rather than a nurse performs triage, are increasingly being used in hospital emergency departments (EDs) to improve patient flow. Before deciding whether such interventions should be adopted, it is important to understand how split flows causally impact patient flow and outcomes.  We employ causal inference methodology to estimate average causal effects of a split flow model on time to be roomed, time to disposition after being roomed, admission decisions, and ED revisits at a large tertiary teaching hospital that uses a split flow model during certain hours each day.  We propose a regression discontinuity (RD) design to identify average causal effects, which we formalize with causal diagrams. Using electronic health records data (n = 21,570), we estimate that split flow increases average time to be roomed by about 4.6 minutes (95\% CI: [2.9, 6.2] minutes) but decreases average time to disposition by 14.4 minutes (95\% CI: [4.1, 24.7] minutes), leading to an overall reduction in length of stay. Split flow is also found to decrease admission rates by 5.9\% (95\% CI: [2.3\%, 9.4\%]) but not at the expense of a significant change in revisit rates. Lastly, we find that the split flow model is especially effective at reducing length of stay during low congestion levels, which mediation analysis partly attributes to early task initiation by the physician assigned to triage.
\end{abstract}

\keywords{Split Flow Model; Causal Inference; Emergency Department; Patient Discharge; Electronic Health Records}

\section{Introduction} \label{sec:intro}
 Americans increasingly receive acute, unscheduled care in the Emergency Department (ED) \citep{pitts2010americans}. ED crowding is an unfortunate result of rising practice intensity and inpatient boarding \citep{pitts2012national,fatovich2005access,olshaker2006emergency,derlet2008ten}. Crowding leads to a number of deleterious effects on patient care including treatment delays, increasing patient mortality, and increased admission rates \citep{sills2011emergency,liu2003impact,mccarthy2009crowding,de2013does,richardson2006increase,sprivulis2006association,mccusker2014increases,gorski2017impact}.  A large number of interventions have been proposed to improve patient flow and thus alleviate crowding, and otherwise improve operational metrics \citep{de2018interventions}. 

One suite of interventions that has gained popularity is the creation of ``split flow" models, in which a Physician or Advanced Practice Provider (APP), rather than a nurse, is stationed at patient intake and briefly sees all walk-in patients. This provider initiates the care of all patients by placing lab, imaging, and medication orders. The provider then stratifies patients, keeping those who do not require a traditional ED bed in a fast track or similar area and moving the rest to a queue for a traditional bed \citep{wiler2016implementation}. Therefore, a split flow model taps into two operational strategies for speeding up care that have been explored by the operations research (OR) and management science (MS) community: (i) physician-triage, whereby an APP initiates tasks that are normally handled later by providers downstream, and (ii) fast-tracking, whereby an APP identifies patients of low acuity and complexity to treat quickly and with fewer resources \citep{batt2017early,freeman2017gatekeepers,russ2010placing}. Efforts to evaluate split flow models have focused on measuring their association with improvement in operational metrics including flow and length of stay, as well as quality and safety metrics \citep{konrad2013modeling,wiler2016implementation,garrett2018effect,pierce2016split,arya2013decreasing,wallingford2018introduction,patterson2020split}. Yet, these associations, while suggestive, do not definitely demonstrate that the split flow model had a direct, or \emph{causal}, effect on outcomes.  Moreover, these pathways may be less effective in reducing patient flow in less crowded EDs and may have adverse consequences downstream from the ED.

Causal effects of split flow models are difficult to measure, since patients were not randomly assigned to split flow model vs. traditional nurse-led triage. With only observational data, average outcomes may be influenced by confounding variables, i.e. factors that influence intervention groups and outcomes. Confounding can lead to spurious associations between interventions and outcomes that are misattributed to the intervention rather than underlying differences in intervention groups. For example, if younger patients are both quicker to treat and more likely to be assigned a split flow model than their older counterparts, then a split flow model may have faster average time to disposition after being roomed than a traditional patient flow model without directly impacting time to disposition. Causal inference methods address this limitation by using conceptual knowledge of system of interest to adjust for possible differences in intervention groups. Examples include direct standardization, which estimates effects within sub-populations, and inverse probability weighting, which estimates effects after re-weighting observations. 

The present study seeks to use causal inference methods to estimate average causal effects of a split flow model on time to be roomed, time to disposition decision after being roomed, admission decisions, and ED revisits.  We use electronic health records (EHR) data on patient visits (n = 21,570) to the ED at a large tertiary teaching hospital. We hypothesize the following:
\begin{enumerate}
\item A split flow model reduces the average time from patient arrival to admission decision.
\item A split flow model has negligible consequences to admission or revisit rates.
\item Faster treatment from a split flow model is mediated by earlier physician orders and by changes in how quickly patients are moved to a bed or vertical patient area. 
\item A split flow model is less effective at reducing average time from arrival to admission decision when the ED is busy. 
\end{enumerate}
In short, our goal is to evaluate the \emph{causal} impact of the split flow model (as opposed to the more traditional nurse-led triage) on \emph{both} operational outcomes (i.e.,time to be roomed, time to disposition decision after being roomed) and patient outcomes (i.e., admission rates, ED revisits). We contribute a causal inference framework that adjusts for measured and unmeasured confounding to evaluate average treatment effects of split flow model on patient flow and patient outcomes.

At the study hospital, the split flow model only operates during certain time periods (i.e., from noon to 9PM) each day. However, patients that arrive to be triaged during operational hours of the split flow model could be different than patients that arrive at other times with respect to age, gender, or unmeasured patient severity and complexity. To overcome this issue, we implement an identification approach for estimating the causal impact of the split flow models on patient outcomes.  This approach relies on a regression discontinuity (RD) design \citep{imbens2008regression,lee2010regression}, which we formalize in a causal diagram. Special consideration is needed to account for an intervention (i.e. split flow model) that is implemented over a specific period of time each day of the week, since ED variables tend to exhibit temporal correlation \citep{whitt2019forecasting}. We thus propose placing a sharp RD design within a linear mixed effects regression model that includes day of arrival as a random effect. We then estimate causal effects using this approach and then compare results.

The remainder of this paper is organized as follows. First, selected studies from the literature are discussed in Section~\ref{sec:related}. The contributions of this paper relative to related topical areas within the literature are highlighted. Next, in Section~\ref{sec:causal}, we formally define a causal framework for evaluating a split flow model. In Section~\ref{sec:identification}, we use this causal framework to present strategies based on a RD method for estimating average causal effects. Upon applying these methods to EHR from a large tertiary hospital, results are presented in Section~\ref{sec:results}. Implications and concluding remarks are given in Section~\ref{sec:conclusion}.

\section{Literature Review} \label{sec:related}

We position our work on split flow models within several areas of the literature. To motivate a split flow model and identify potential reasons why this model may improve patient flow, we consider work in the OR/MS community that investigate strategies involving an APP for improving ED patient flow. We then consider simulation and queueing approaches of split flow models, since the present work should inform what parameters are realistic for these approaches. Next, we consider work in the clinical community that uses observational data to estimate the impact of having an APP on patient outcomes, since we will similarly take an empirical approach. Last, we discuss causal inference methods, with a focus on RD design, to motivate our approach to estimating the direct impact of split flow models on outcomes.

\subsection{Operational strategies to improve ED patient flow}
Optimizing patient flow in the ED has generated great interest from the OR/MS community, given the difficulty EDs face across the US to address overcrowding and long waiting times \citep{saghafian2015operations}. With constraints on resources in the ED, these efforts focus on how one might reorganize these limited resources for diagnosing, monitoring, and treating patients. One such reorganization is \emph{physician-triage}: the nurse at triage is replaced with a provider who would have typically seen the patient downstream \citep{russ2010placing,partovi2001faculty,traub2016physician}. Physician-triage may improve patient flow by initiating tasks, such as diagnostic labs and images, earlier in an ED visit \citep{batt2017early}. This may allow providers to wait less for results from these tests they need to decide upon suitable treatment and whether to admit the patient. Another reason physician-triage may improve patient flow is providing a more precise assessment of patient needs or patient complexity \citep{saghafian2014complexity,saghafian2018workload}. Physician-triage may be thought of as a form of gatekeeping \citep{freeman2017gatekeepers,freeman2020gatekeeping}, where one provider determines who receives additional services, i.e. acts as a ``gatekeeper" of additional services. 

Another reorganization strategy is \emph{patient streaming}: where patients are streamed into different physical locations in the ED based on their needs \citep{saghafian2012patient}. Here, the idea is to more optimally match resources to patient needs, so that patients do not receive unnecessary resources that could have gone to elsewhere. Fast-tracking is one example of patient streaming, which specifically targets low-needs patients who do not require an ED bed to be treated \citep{considine2008effect}. 

A split flow model leverages both reorganization strategies (i.e. physician-triage and fast-tracking) for reducing waiting times. This work seeks to measure the direct contribution of this combined strategy to patient outcomes and explore the degree to which any direct effect might be mediated by early task initiation or fast-tracking. Furthermore, given that providers can adapt to the working context in the ED even without the reorganization \citep{batt2017early,gorski2017impact}, it is important to understand how the effectiveness of a split-flow model might change with increasing workload and overall congestion.

\subsection{Stylized split flow models}

With a growing number of EDs adapting split flow models, the OR and MS community have used simulation and queueing models to explore the impact of split flow models on patient flow and costs (c.f., \citet{konrad2013modeling,zayas2016dynamic,zayas2019policies,kamali2018use} and references therein). For example, \citet{zayas2016dynamic} and \citet{zayas2019policies} studied how providers should prioritize their work in a split flow model to balance initial delays for care at triage with the need to discharge patients in a timely fashion. Triage and treatment were modeled as a two-stage tandem queue. Allocation policies were analyzed using continuous-time Markov decision processes (CTMDPs) \citep{zayas2016dynamic} and simulation \citep{zayas2019policies}.  \citet{kamali2018use} studied when a physician should lead triage over a nurse in an ED using a fluid approximation of a two-stage queueing model and simulation. Each of these models use data from a partner hospital to recover realistic modeling parameters, but evaluate policies in terms of objectives (e.g., abandonment costs) that might reflect financial considerations, patient flow, and/or patient outcomes. These queueing models may benefit from objectives that account for the actual impact of the split flow model on patient outcomes, which the present paper seeks to estimate. 

\subsection{Actual split flow models}

Several empirical studies have analyzed whether split flow models are associated with decreased length of stay (LOS) (c.f., \citet{subash2004team,medeiros2008improving,han2010effect,Soremkun12}) or reduced ED revisits and mortality (c.f.,  \citet{Burstrom12,burstrom2016improved}). For example, in a retrospective study of three Swedish EDs, \citet{Burstrom12} consider three triage models: senior physician-lead team; nurse first, emergency physician second; and nurse first, junior physician second. They show that physician-led triage is associated with reductions in the time to first doctor encounter, LOS, rate at which patients leave without being seen, revisits, and mortality. These associations suggest split flow models improve patient flow and outcomes, but since the intervention was not randomized, other factors that differ between the three EDs could have driven the discrepancy in patient flow and outcomes. 

Empirical support has also been provided for certain operational strategies (e.g., physician-triage, fast-tracking) embedded in the split flow model. \cite{russ2010placing} found that patients who had an order placed by a physician prior to bed assignment saw a 11-minute decrease in median time until disposition compared to a matched sample of individuals with a similar order placed after bed assignment. Using data from 16 consecutive Mondays, \cite{partovi2001faculty} found a decrease in LOS on the 8 Mondays with physician-triage compared to 8 days without. \cite{traub2016physician} also investigated physician-triage, wherein they used a log-linear regression model to predict log LOS after matching days with a physician triage to days in the next year based on day of the week and calendar date. They found a 6.25\% decrease in the geometric mean LOS due to physician-triage compared to rotational patient assignment based on a linear regression model. \cite{considine2008effect} found fast-tracked patients saw a 16-minute decrease in median LOS compared to matched controls from a prior time period when the ED did not use fast-tracking. Critically, all these studies use a matching strategy, which is useful for adjusting for measured confounding, e.g., the day of the week that a patient arrives, but not unmeasured confounding (e.g., patient severity). Perhaps as a way to adjust for unmeasured confounding, some of these studies also adjust for variables that could be affected by the intervention itself (e.g., disposition disposition, patients registered in a day). However, these adjustments can make it difficult to interpret the estimate of interest as an average causal effect.

\subsection{Causal inference methods}

In an effort to go beyond associations, causal inference methods estimate the average causal effects of an intervention, relying on \emph{identification} assumptions to ensure estimates are unbiased or consistent. Among the various causal inference methods available, we focus on a regression discontinuity (RD) design for its potential for adjusting for both measured and unmeasured confounding. 

Originally considered by Thistlethwaite and Campbell \citep{thistlethwaite1960regression}, RD designs were developed to estimate the average impact of interventions in non-experimental settings where the intervention is determined by whether an observed variable, referred to as the ``forcing" or ``running" variable, exceeds a known cutoff point. A RD design fits our problem: we want to estimate the average impact of the patient-flow model (split flow vs. traditional flow) and can use arrival time as the forcing variable, since the split flow model operates during certain hours of the day. Broadly, the idea of RD design is to build a regression model to explain outcomes as a function of the forcing variable using data only with a forcing variable near the cutoff point. Under certain assumptions, \emph{discontinuity} in mean outcomes at the cutoff point can then be attributed to the intervention \citep{hahn2001identification}. 

Many studies have relied on RD designs to estimate average causal effects of interventions.  We refer the reader to the user guides by \citet{lee2010regression} and \citet{imbens2008regression} for a discussion of practical and theoretical considerations of RD methods.  The study by  \citet{almond2011after} is closest to ours in terms of application area and using time of day as a forcing variable.  They use a RD design to study extending the hospital stay of newborn affects costs and outcomes. Given that newborns with longer hospital stays may have different insurance coverage and treatment needs, \citet{almond2011after} compared newborns born right before and right after midnight, since newborns after midnight were reimbursed for an extra day in the hospital compared to newborns before midnight. They find that the additional reimbursed day induces substantial extensions in length of hospital stay for mother and their newborn, but did not have a significant effect on readmissions or mortality. Although similar to the approach of \citet{almond2011after}, our approach will have an important difference. Specifically, we account for potential non-independence of observations from strong intra-day correlation using a mixed effects regression model.

\section{Causal Framework} \label{sec:causal}

\smallskip

\subsection{The intervention: a split flow model for the ED}

\begin{figure}
    \centering
    \includegraphics[width=\textwidth]{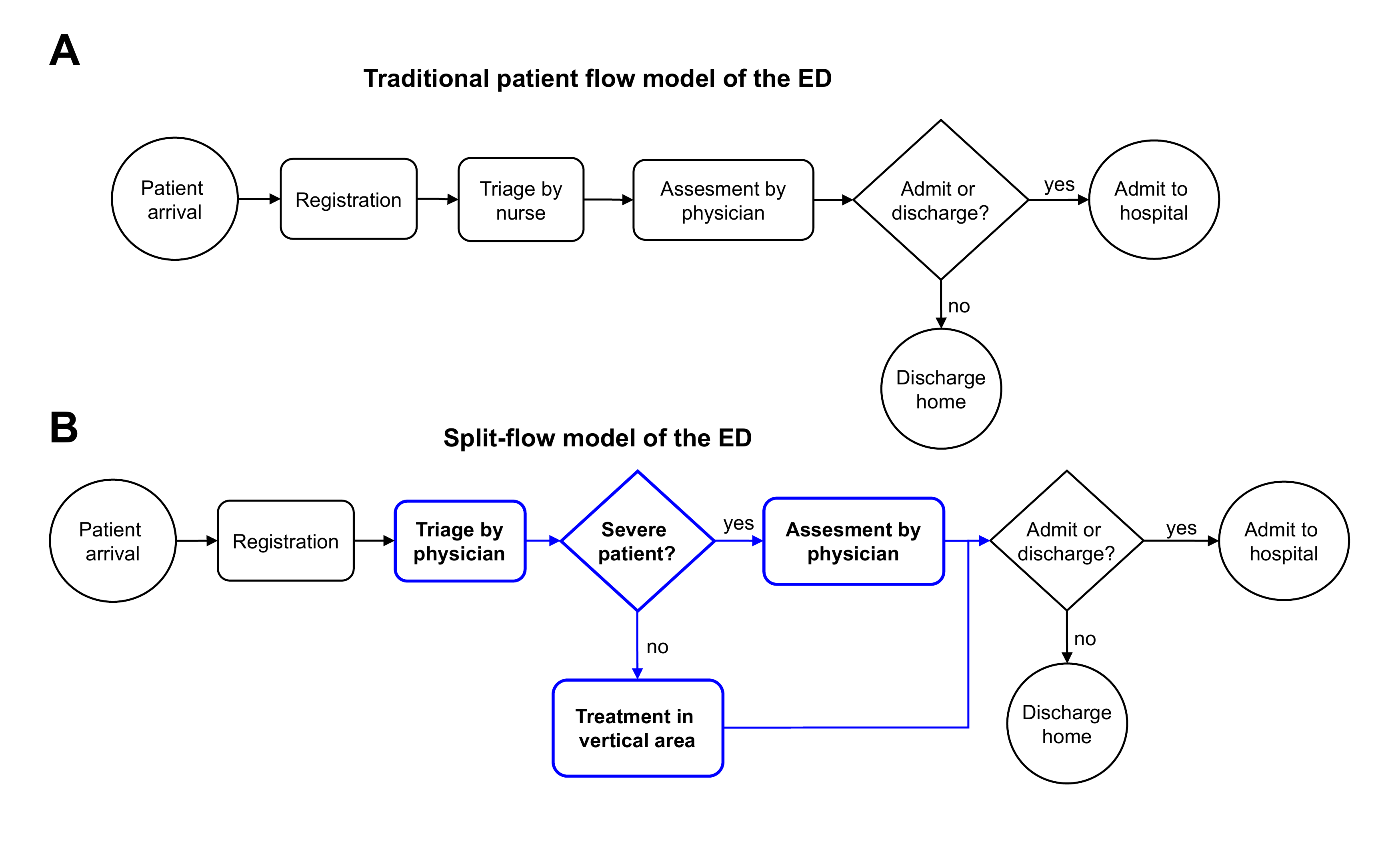}
    \caption{Patient flow models of the ED. \textbf{(A)} In a more traditional patient flow model, a nurse triages patients after which a physician assesses and treats the patient. \textbf{(B)} In a split flow model, a physician triages a patient which allows them to immediately treat the more severe patients.}
    \label{fig:patient_flow}
\end{figure}

Although variations occur depending on the needs of the patients and the specific hospital under consideration, most visits start with patients registering at a check-in desk and recording a chief complaint (e.g. abdominal pain, chest pain, shortness of breath, headache).  After registration, a patient can follow a more traditional model of patient flow (Figure ~\ref{fig:patient_flow}A) or a split flow model (Figure ~\ref{fig:patient_flow}B). 

In traditional patient flow, the patient is triaged by a nurse during which information is collected about their vital signs and why they came to the ED.  After triage, the nurse assigns patients an acuity (ESI or emergency severity index) score of 1-5. These scores along with patient complaints determine the order in which the patient will be brought into a main assessment/treatment room.  If they are not immediately brought back to a room, they remain in the waiting room. When it is their time, the patient is placed in a room where a provider in the ED will assign themselves to the patient, initiating the physician or physician assistant (PA) interaction with the patient.  After assigning themselves, physicians will see the patient, enter orders for treatments and tests including labs and imaging.  Patients will receive care from nurses and physician, and will often need reassessment to ensure they are responding appropriately prior to disposition.  Other providers from services like orthopedics, neurology, etc., may be called by ED staff to see the patient at some point after providers are assigned.  After treatment, a determination of where a patient will go (i.e., transfer, discharge, or admission) is made.

By contrast in a split flow model, the role of the nurse responsible for triage is replaced by an attending physician focused on starting care, developing care plans, and sorting the patients in the correct queue for care.  At our partner ED, this process started at 1pm to 10pm from November 1, 2016 to June 29th, 2017 and henceforth starts at 12pm and lasts until 9pm each day and during those times, and takes place in a dedicated suite of rooms immediately after the patient arrives. After triage, patients who require a bed in the main ED are sent to a bed in the main ED. There, they will wait for any care initiated by the physician at triage or for care from the physicians devoted to the main ED. Lower acuity patients are discharged if they do not need specialized attention or guided through specially designated care area if the triage physician decides that they will require additional medical care. The designated rooms are cleaned and prepared for the next patient. 
Patients assigned to these rooms will wait for care from the physician responsible for triage or the definitive care provided by the care team, or for medical treatments such as radiology or IV placement. Finally, after the treatment inside this room ends, a decision of discharge or admission is made.

To summarize, a split flow model, like many interventions, consists of several key components, each of which are inherent to the intervention and could contribute to effectiveness over a traditional patient flow. These include
\begin{itemize}
    \item Initial triage and early task initiation by a physician
    \item Treatment in a vertical patient (no bed) area of patients who do not require beds
\end{itemize}
As such, we are investigating the sum of these components. Each was designed to improve patient flow. As a result, we expected:
\begin{hyp}
A split flow model reduces the average time from patient arrival to admission decision. 
\end{hyp}

\smallskip
\noindent Each component was designed to re-organize patient flow, not patient care. So, we had also expected that:
\begin{hyp}
A split flow model has negligible consequences to admission or revisit rates.
\end{hyp}

\smallskip
\noindent Moreover, any speed up may be attributed to either component of a split flow model: early task initiation or streaming of patients to a bed or the vertical patient area. This motivates our next hypothesis:
\begin{hyp}
\noindent Faster treatment from a split flow model is mediated by earlier physician orders and by changes in how quickly patients are moved to a bed or vertical patient area. \end{hyp}

\smallskip
Finally, ED providers speed up care when the ED is busy even without a physician at triage, which could limit the effectiveness of a split flow model when the ED is busy. So, we also expected that:
\begin{hyp}
A split flow model is less effective at reducing average time from arrival to admission decision when the ED is busy. \end{hyp}

\smallskip
\noindent We now define the necessary variables to be able to test these hypotheses. 

\subsection{Average causal effect of the intervention}

We use the potential outcome framework, central to the counterfactual theory of causality proposed by Neyman \citep{neyman1923application} and extended by Rubin \citep{rubin1974estimating} and Robins \citep{robins1986new}, to examine how the intervention, a split flow model, directly impacts four common measures of ED care: time to be roomed, time to disposition decision after being roomed, admission decision, and ED revisits. 
For example, ED revisits is a common metric for evaluating ED care since it might indicate that follow-up care was not properly organized or that patients were not adequately treated before discharge \citep{keith1989emergency,sabbatini2016hospital}. However, we caution the reader that this metric has come under recent scrutiny as a measure of quality care \citep{welch2009quality,rising2015return,cheng2016emergency,shy2018bouncing}. Letting $Y \in \R$ denote observed outcomes, potential outcomes $Y^a \in \R$ denotes the random potential outcome that could have been observed were a split flow model used ($a=1$) or were a more traditional patient flow model ($a=0$). Within this framework, we can examine the direct impact of the split flow model on potential outcomes by examining how potential outcomes differ on average between the two patient flow models: $\E[Y^1 - Y^0]$. 

Because we cannot observe both potential outcomes for each visit, we will need additional variables to identify potential outcomes. Patient arrival time is denoted by $T$ with units of hours. Patient flow model $A \in \{0,1\}$ is a deterministic function of $T$, with $A=1$ signifying that a split flow model was used and $A=0$ signifying a traditional flow model. Patient arrival time may depend on measured characteristics $X$ prior to treatment (e.g., age, sex) as well as unmeasured variables  $U$. For example, patients with less urgent conditions often wait until after work hours to visit the ED or use the ED on the weekend when they cannot reach their primary care provider \citep{davis2010identifying}. Further, all variables might influence patient outcomes. 

\subsection{Challenges with identifying causal effects}

Identifying average causal effect $\E[Y^1 - Y^0]$ requires three assumptions: consistency, positivity, and exchangeability \citep{hernan2010causal}. Consistency holds if the intervention is well-defined in that the actual outcome $Y$ is $Y^A$, i.e. the potential outcome under the observed patient flow model $A$. We assume consistency of the patient flow model. Positivity holds if every patient has a chance of being assigned to either patient flow model: $0 < \PR[A=1] < 1$. Exchangeability holds if potential outcomes are independent from the patient flow model: $Y^0,Y^1 \ind  A$. Broadly speaking, this means that actual outcomes are representative of potential outcomes. In practice, positivity and exchangeability may not be reasonable for the entire sample, but can be weakened to hold only within groups of individuals with similar characteristics.  

\begin{figure}[ht!]
    \centering
    \includegraphics[width=0.95\textwidth]{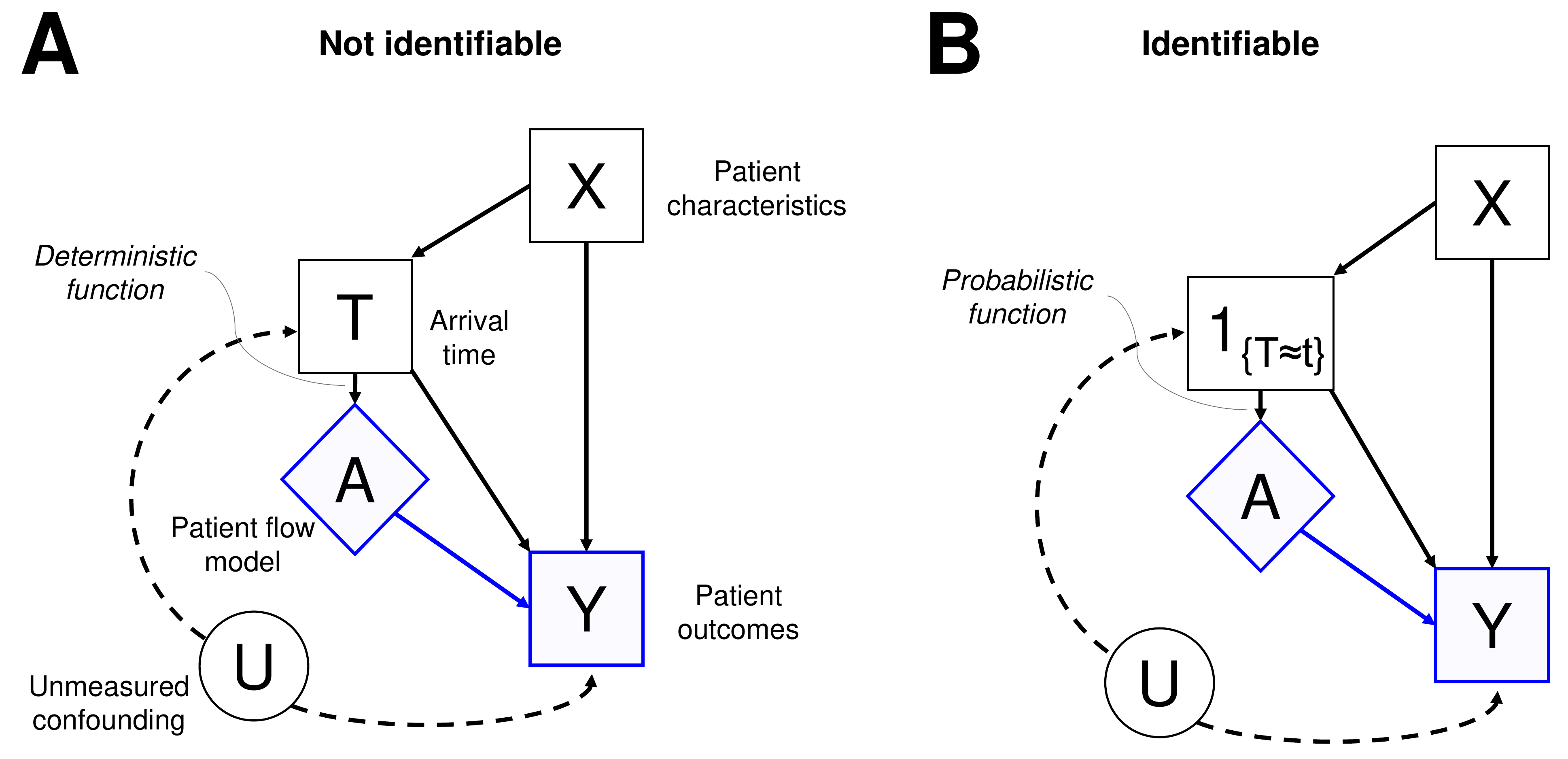}
    \caption{Causal diagrams of patient outcomes. Causal effects of patient flow model $A$ on patient outcomes $Y$ (in blue) are generally not identifiable in the diagram in \textbf{(A)}, as either positivity or exchangeability is violated, but could be identified under additional assumptions as specified in the diagram in \textbf{(B)}.}
    \label{fig:DAG}
\end{figure}

We anticipate that either positivity or exchangeability might not hold based on our conceptual model of the ED, which we depict in the causal diagram in Figure~\ref{fig:DAG}A. This diagram captures the fact that arrival time $T$ determines the patient flow model $A$ and the possibility that the patient flow model $A$ influences outcomes, since this relationship is our question of interest. It also includes the possibility that patient characteristics $X$ directly influence the arrival time $T$ and that patient characteristics $X$ and arrival time $T$ also directly influence outcomes $Y$. For example, younger patients ($<$ 65 years) may be more likely to arrive outside work hours, or hospitals may want to speed up time to disposition during peak ED hours or slow down time to disposition for their older patients. Finally, we anticipate possible unmeasured variables $U$ which could directly influence both arrival time $T$ and patient outcomes $Y$. Unmeasured patient severity, for instance, might lead to worse outcomes and an arrival time $T$ in early hours. 

In general, average causal effects are not identifiable when the relationship between the patient flow model $A$ and patient outcomes $Y$ is indeed confounded by unmeasured variables $U$. Based on Figure~\ref{fig:DAG}A, patients with the same arrival time $T$ are exchangeable, i.e. $Y^0,Y^1 \ind \, A \, | \, T=t\,$. Exchangeability is seen for this diagram as the blocking of any backdoor paths from $A$ to $Y$ by the variable $T$. Patient with the same arrival time $T$, however, do not have a positive probability of being assigned to either patient flow model, i.e. $\PR[A=1|T=t] \in \{0,1\}\,$, since the patient flow model is a deterministic function of $T$. On the other hand, patients with the same patient characteristics $X$ have a positive probability of being assigned to either patient flow model, i.e. $0 < \PR[A=1|X=x] < 1$, but are generally not exchangeable $Y^0,Y^1 \notind A | X=x$, since $X$ does not block backdoor paths from $A$ to $Y$ that go through $T$.

Causal effects could be identified given the diagram in Figure~\ref{fig:DAG}B. This diagram can identify effects provided a coarser variable $1_{\{T \approx t\}}$ of arrival time can guarantee positivity and maintain exchangeability for patients with similar values of $1_{\{T \approx t\}}$. These observations will motivate our approach for estimating $\E[Y^1 - Y^0]$.

\subsection{Data available for identification}

Visits to the study ED (n = 112,083) were available to evaluate the intervention. Visits occurred between November 1st, 2016 to September 27th, 2018. Data was curated from EHR of the study hospital and an associated network of primary and specialty care clinics. Visits were excluded sequentially for the following reasons: patients were transferred from other hospitals or hospices, since these patients were triaged outside of the study ED (n = 1,979); chief complaint was misreported as a procedure (n = 143); patients were younger than 18 years, since they were never triaged under a split flow model at the study ED (n = 23,664); patients arrived by ambulance, helicopter, or other means different than walking through the registration process of the ED (n = 25,886); patients had a missing time to disposition decision after being roomed (n = 26); and patients had a missing disposition decision (n = 20).  The remaining sample 
was used for analysis. 

The sample contained baseline patient characteristics $X$: \ul{age}, \ul{sex}, \ul{race/ethnicity}, \ul{chief complaint}, and \ul{insurance}. Observations were also available for \ul{arrival time} $T$ and the \ul{split flow model} $A$ (a binary outcome [yes/no] specifying whether patient utilized the split flow model) and for each of four outcomes $Y$, analyzed separately: 
\begin{itemize}
\item \underline{Time to be roomed}:  duration of time between arrival time and when patient is placed in a treatment or vertical area room,
\item \underline{Time to disposition}: duration between when a patient is roomed and when the disposition decision is made,
\item \underline{Admission decision}: a binary variable [admit/discharge] specifying the decision to discharge patient home or admit to them to inpatient unit,
\item \underline{ED Revisit}: a binary outcome [yes/no] specifying whether the patient returns to the ED within $30$ days of being discharged from either the hospital for admitted patients or the ED for discharged patients.
\end{itemize}
To simplify analysis, we grouped chief complaint into the five most common ones (i.e., abdominal pain, chest pain, dyspnea, fall, and fever) and marked the remaining complaints as ``Other". A missing chief complaint (n = 151) was also marked as ``Other". Visits with missing health insurance (n = 9,355) were placed into the ``Unknown"  category.

\subsection{Confounding, moderation, and mediation}
Our main explanatory variable $A$ is whether an ED patient goes through a split flow model or a traditional patient flow model. However, several variables may offer further operational insight into the implementation of a split flow. We organize these variables within our causal framework into confounding, mediating, or moderating variables.

\subsubsection{Confounding variables.}
Our ability to estimate the average causal effect of the split flow model, i.e. the specific intervention that involves a physician at triage and a vertical area dedicated to treating low acuity patients, will rest on the assumption that the arrival time $T$ will block any variable that confounds the relationship between the intervention $A$ and outcome $Y$ (see Figure~\ref{fig:DAG}). This assumption should account for \emph{patient-level} differences between split flow and traditional patient flow patients, provided patients do not manipulate their arrival time. However, this assumption does not account for possible \emph{operation-level} differences between split flow and traditional patient flow patients at our study hospital, which could confound the relationship between the intervention and outcome. To attribute any improvement to solely the split flow model, we must investigate potential operation-level confounders in our particular dataset:
\begin{itemize}
    \item \ul{Shift change}. Physicians are assigned to seven different shifts at the study hospital, one of which is a one-person shift devoted to implementing the split-flow model from 12a--9p. Patient care may improve due to the shift change at the start of the split-flow model rather than the split-flow model itself. If a shift change is important, then we may expect to see an effect on patient outcomes from changes to other shifts (i.e., 7a--4p, 11a-8a, 9a--5p, 5p--2a, 10a--4p, 4p--2a). 
    \item \ul{Physician assignment}. How physicians are scheduled to shifts may introduce bias if certain physicians were both disproportionately assigned to split flow and more likely to produce certain outcomes compared to other physicians. Thus, we consider the particular physician assigned to a patient visit as a possible confounder.
\end{itemize} 

\subsubsection{Moderating variables.} 
The split flow-model may be more or less effective depending on operational context. This information will be useful when trying to predict the degree to which a split flow model will be effective in a particular hospital at various points in time. We consider whether the following operational variables may moderate the relationship between the intervention and patient outcomes:
\begin{itemize}
    \item \ul{Congestion.} One of our main hypotheses is that split flow model may improve patient flow more greatly during periods of high congestion. We computed a variable, which we call congestion, for each visit in our dataset as the ED census at the time of arrival. The ED census included everyone in the adult ED, regardless of who they are or where they are in the ED (e.g., waiting room) or in their treatment pathway (e.g., post-disposition decision). 
    \item \ul{Day of the week.} Given our hypothesis about congestion and how congestion varies over the 7-day week, we also expect that day of the week will be another moderator.
    \item \ul{Physician workload.} Like congestion, physician workload might also be a possible moderator, with higher physician workloads associated with greater improvements from the split-flow model. One benefit of physician workload over congestion is that it accounts for the number of physicians in the ED. We can compute workload for each physician at each point in time following prior literature \citep{batt2017early,kc2014does,song2015diseconomies} as the ratio of patients to providers. To recover a variable, which we will call physician workload, for each visit, we take the ED census at the time of arrival averaged over physicians currently in the ED. We point out that this variable uses only information available prior to being assigned a patient flow model (as opposed to the workload of any physician assigned to the visit), which allows us to treat this variable as a moderator.
    \item \ul{Start time of split flow model.} When the split flow model is initiated during the day may be another moderator. Since the start time for the split flow model changed from 1p to noon in our dataset, we are able to investigate start-time as a possible moderator. 
\end{itemize}

\subsubsection{Mediating variables.}
When describing the split flow, we pointed out two components of the split flow model: initial triage and early task initiation by physician and treatment in a vertical area of low-needs patients. Ideally, we want to learn the degree to which each of these components contributes to intervention effectiveness, so that hospitals could optimize the implementation of split flow. However, we cannot directly separate their contribution, since they are both inherent to the intervention and operate at the same time. We might indirectly get at this information by investigating whether certain markers of these components might mediate the relationship between the intervention and patient outcomes. Hence, we consider the following:
\begin{itemize}
    \item \ul{Time to first order}. We consider the time to first order of either imaging, labs, or medication.  This variable is used as a mediator to capture the hypothesized benefits of early task initiation. In particular, if early task initiation is important to the split-flow model's effectiveness, then we expect to find that a split flow model leads to shorter time to first order of lab tests and that shorter times to first order, regardless of the patient flow model, leads to shorter time to disposition. 
    \item \ul{Time to be roomed.} While this variable is considered a primary outcome, it might also act as a possible mediator of the relationship between a split flow model and time to disposition after being roomed. On the one hand, treatment can be initiated early for low-needs patients by the physician at triage before they are officially placed in the vertical area. As a result, we might expect that split flow might increase time to be roomed and that longer times to be roomed, specifically under a split flow model, lead to shorter time to disposition after being roomed. On the other hand, the physician is able to quickly stream low-needs patients to the vertical area. As a result, we might expect that split flow might decrease time to be roomed, but at the expense of longer times to disposition.
\end{itemize}

A potential outcomes framework can again be used to define different mediation effects \citep{nguyen2020clarifying,imai2010general}. To define these effects, let $M$ be a random variable representing one of the possible mediators, either time to be roomed or time to first order (of either imaging, labs, or medication). Let $M^a \in \R$ denote the random potential mediator that could have been observed were a split flow model used ($a=1$) or a traditional patient flow model ($a=0$). In addition, let $Y^{a,b} \in \R$ to denote the random potential outcome that could have been observed were both a split flow model used ($a=1$) or a traditional patient flow model ($a=0$) and the mediator variable was set to $b$. Similar to potential outcomes, we assume consistency (i.e., $M=M^A$ and $Y=Y^{A,M}$). 

In terms of the variables above, the average causal effect can be decomposed in one of two ways:
\begin{align*}
    \E[Y^1 - Y^0] &= \E[Y^1 - Y^{1,M^0}] + \E[Y^{1,M^0}-Y^{0}], \\
    \E[Y^1 - Y^0] &= \E[Y^1 - Y^{0,M^1}] + \E[Y^{0,M^1}-Y^{0}].
\end{align*}
This decomposition informs us about mediation. There are average \emph{natural indirect} effects (NIE):
\begin{align*}
    \bar{\delta}(a):=\E[Y^{a,M^1} - Y^{a,M^0}]
\end{align*}
for $a=0,1$, capturing the degree to which the effect of an intervention is attributed to changes in the mediator. There are also \emph{natural direct} effects (NDE):
\begin{align*}
    \bar{\zeta}(a):=\E[Y^{1,M^a}-Y^{0,M^a}],
\end{align*}
for $a=0,1$, capturing the degree to which the intervention has an effect were the mediator fixed. In the context of the present study, the NIEs reflect the gap in average outcomes if we were to change the time to be roomed or time to first order under a split-flow vs. patient flow model but keep the rest of the patient flow model fixed. The NDEs reflects the gap in average outcomes between those who undergo a split flow model vs. a traditional flow model while keeping the time to be roomed or time to first order fixed. A possible mediator is important to a split flow model if we observe large NIEs but small NDEs. 

\begin{figure}[ht!]
    \centering
    \includegraphics[width=0.45\textwidth]{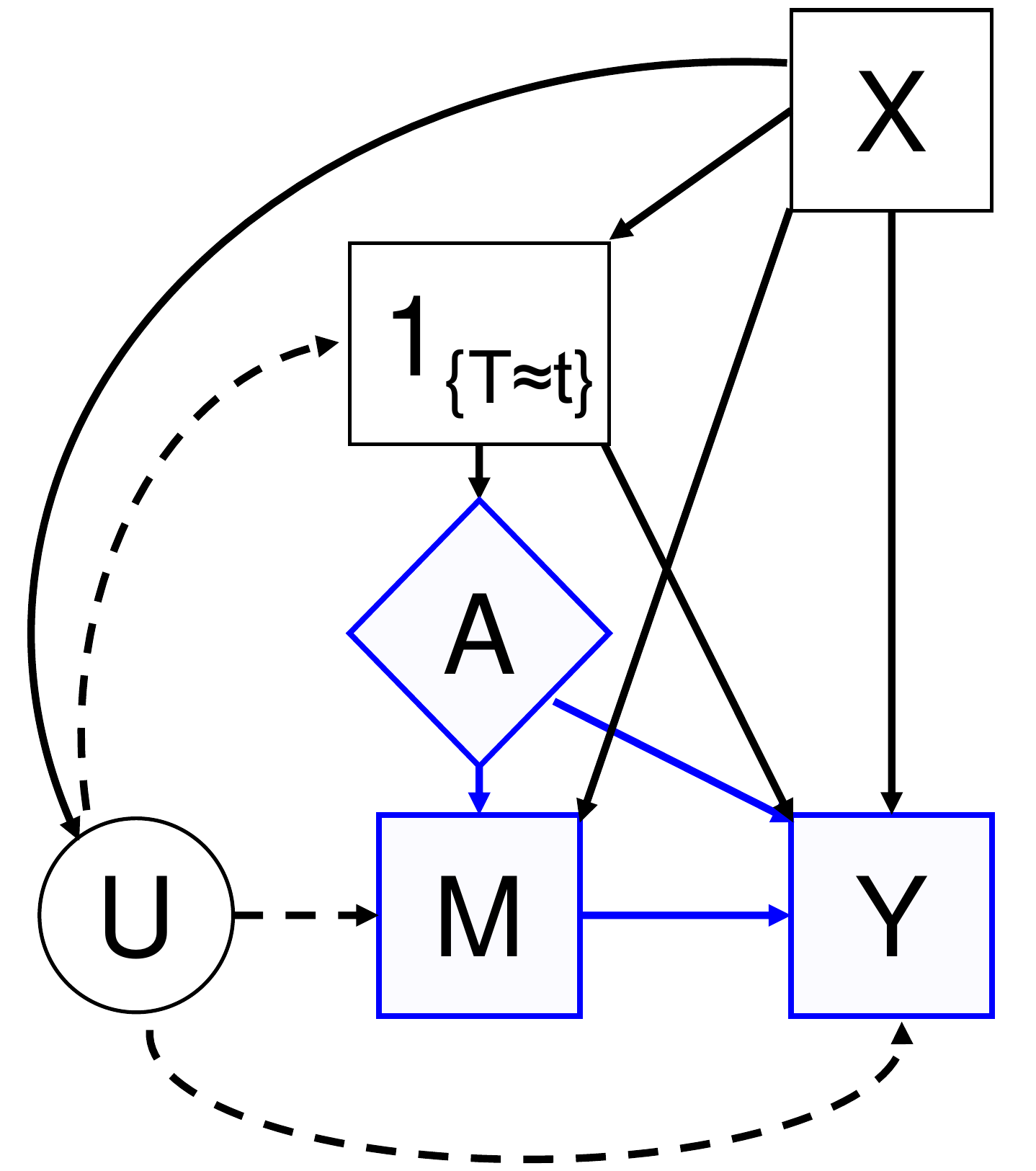}
    \caption{Causal diagrams of patient outcomes including a mediator $M$. Causal effects of patient flow model $A$ on patient outcomes $Y$ which involve a mediator $M$ (in blue) are generally not identifiable in the diagram, as exchangeability is violated.}
    \label{fig:DAG_Mediation}
\end{figure}

Identifying NIEs and NDEs imposes additional challenges beyond what is faced when trying to identify the average causal effect. Even though arrival time $T$ may block backdoor paths through $A$, we do not expect arrival time $T$ to block backdoor paths through the mediator $M$ (Figure~\ref{fig:DAG_Mediation}). Rather, both measured variables $X$ and unmeasured variables $U$, notably patient severity, are expected to influence both the mediator $M$ (i.e., whether an individual is treated in the vertical area or receives an early image order) and patient outcomes $Y$. This means that exchangeability is violated, and the NIEs and NDEs are not identifiable in general. In such a case, specific assumptions about the relationship between variables are needed for identification.


\section{Identification of average causal effects}\label{sec:identification}

\smallskip

\subsection{Sharp regression discontinuity}

To identify average causal effects, we consider a sharp regression discontinuity (RD) by restricting attention to patients with arrival times $T$ around the time each day that the split flow model is started. Intuitively, we want patients with similar arrival times ($T\approx t$) to be exchangeable (as in the identifiable causal diagram in Figure~\ref{fig:DAG}B). Consequently, any difference in potential outcomes around the start times of a split flow model can be attributed to the patient-flow model. Formally, a sharp RD design requires that there exists some time point $t_0$ at which the probability of patient-flow model is discontinuous in arrival time $T$  (Assumption~\ref{ass:rd1}) but potential outcomes are continuous in $T$ (Assumption~\ref{ass:rd2}): 

\begin{assumption}{(Discontinuity of treatment assignment)}\label{ass:rd1}
Limits $\lim_{t \to t_0^{+}} \, \PR[A=1 | T = t]$ and $\lim_{t \to t_0^{-}} \PR[A=1 | T = t]$ exist but are unequal. 
\end{assumption}

\begin{assumption}{(Continuity of potential outcomes)}\label{ass:rd2}
 $\E[Y^a | T=t]$ is continuous at $t=t_0$ for $a=0,1$.
\end{assumption}

Under these two assumptions, average causal effects can be identified at $t=t_0$ \citep{hahn2001identification}, since
\begin{align*}
\E[ Y^1 - Y^0 | T=t_0 ] = \lim_{t \to t_0^{+}} \E[Y | T = t] - \lim_{t \to t_0^{-}} \E[Y | T = t]. 
\end{align*}
For our problem, Assumption~\ref{ass:rd1} is trivially satisfied at every start time of the split flow model by definition of the patient flow model $A$.  Assumption~\ref{ass:rd2} will hold if individuals are exchangeable around start times of the split flow model, but is generally not verifiable. There are, however, techniques to evaluate this assumption as we discuss in the next subsection. 

\begin{remark}
Note there is no theoretical reason why we could not also consider times $t_0$ each day when the split flow model \emph{ends}. Practically, however, there are fewer arrivals around the split flow model's daily end time that are available for estimation. An average of 11 patients arrive between 11a and 1p (12a and 2p before November 1, 2017) compared to only 8 between 8p and 10p (9p and 11p before November 1, 2017) each day. For simplicity, we thus focus on using a RD approach for start times of the split flow model in the main text, leaving an analysis of end times in the Appendix.
\end{remark}

While average causal effects can be identified with these assumptions for any start time $t_0$ of the split flow model on a given day, there are few visits around each start time $t_0$ available for estimation. Fortunately, whether or not a patient is assigned to a split flow model depends only on the time of day that they arrive (e.g., 1:05p) and not the exact date and time that they arrive (e.g., 12/01/20 at 1:05p). Thus, we suspect that patients are exchangeable immediately before or after the start time of the split flow model. That is, letting $S$ denote the difference between the arrival time $T$ and the start time on the given day modulo 24 hours, we believe patients are exchangeable if $S\approx 0$. In light of these observations, we replace Assumptions~\ref{ass:rd1}--\ref{ass:rd2} with
\begin{assumption}\label{ass:agg_rd1}
Limits $\lim_{s \to 0^{+}} \, \PR[A=1 | S = s]$ and $\lim_{s \to 0^{-}} \PR[A=1 | S = s]$ exist but are unequal. 
\end{assumption}
\begin{assumption}\label{ass:agg_rd2}
 $\E[Y^a | S=s]$ is continuous at $s=0$ for $a=0,1$.
\end{assumption}
Under these two assumptions, we can then identify the average causal effect of the patient flow model at $s=0$ aggregating data from every day that the split flow model is implemented.

\subsection{Estimation of average causal effects}
We estimate \texorpdfstring{$\E[ Y^1 - Y^0 | S=0 ]$}{E[Y1-Y0|S=0]} by assuming certain structural dependencies between time from arrival to split flow start $S$ and outcome $Y$. Specifically for continuous outcomes, we assume outcomes satisfy: 
\begin{align*}
\E[Y | S ] =  \alpha + \begin{cases} f_0(S) & \text{when } S < 0 \\
f_1(S) + \gamma & \text{when } S \geq 0
\end{cases}
\end{align*}
\noindent for some polynomial functions $f_0(\cdot)$ and $f_1(\cdot)$ with $f_0(0)=f_1(0)=0$. It is common to use low degree polynomials for $f_0$ and $f_1$, such as linear or quadratic functions, since a high degree polynomial can lead to misleading estimates \citep{gelman2019high}. These functions can also be constrained to share coefficients such as slopes. Provided the model is correctly specified and under Assumptions~\ref{ass:rd1}--\ref{ass:rd2}, then taking left and right limits towards $s=0$ reveals that the coefficient $\gamma$ is  exactly the desired average causal effect $\E[ Y^1 - Y^0 | S=0 ]$.

Although average causal effects are identified with only a structural model of $Y$ in $T$, regression models can usually increase efficiency in estimation by adding covariates to explain additional variation in $Y$. We thus added baseline covariates $X$ to the regression model:
\begin{align}\label{eq:rho_model}
\E[Y | S, X ] = \alpha + \beta X + 
\begin{cases} f_0(S) & \text{when } S < 0 \\
f_1(S) + \gamma & \text{when } S \geq 0
\end{cases}.
\end{align}
Specifically, we adjusted for age, sex, race (White vs. Non-White), and chief complaint (abdominal pain vs. other). To estimate these parameters, we note that observations collected on any given day may be strongly correlated, given that ED variables tend to exhibit significant correlation patterns, e.g., daily, weekly, and monthly \citep{whitt2019forecasting}. We thus adjusted for this non-independence by estimating parameters with a linear mixed effects regression model in which a random intercept is included for each day of arrival. Although both a random intercept and a fixed intercept can account for non-independence, we opted for a random intercept to dramatically reduce the number of parameters estimated: 2 for the mean and standard deviation of the random intercept vs. 696 for the fixed intercepts associated with each day in the sample. An alternative is to estimate parameters for each day in the dataset and then aggregate these estimates.

For the two binary outcomes (admission and ED revisit), we use the same model at \ref{eq:rho_model} to estimate $\gamma$. In this case, the model is referred to as a \emph{linear probability model}, and $\gamma$ reflects that the difference in admission/revisit rate resulting from the split flow model. An alternative approach would be to apply a logistic (or probit) transformation to $\E[Y|T,X]$, where then $\gamma$ would have been the limiting odds ratio between treatment and control groups. Unfortunately, even when covariates $X$ and intervention $A$ are all mutually independent, logistic or probit regression models are not \emph{collapsible}, as defined in \cite{greenland1999confounding}, whereas linear or log-linear regression models are collapsible. Noncollapsibility means that the intervention effect $\gamma$ as a measure of association between $A$ and $Y$ changes with the choice of covariates $X$ and can occur even when the intervention effect is equal in every level of a covariate. Since we consider regression models with different covariates, we wanted to avoid the difficulty of comparing and interpreting an intervention effect that is particularly sensitive to our choice in covariates. One caveat of using linear probability models for probabilities is that they are not constrained to lie in the interval $[0,1]$, leading to biased estimates that depend on the number of observations for which the fitted model yields predicted outcomes outside the interval $(0,1)$ \citep{horrace2006results}.

\subsection{Graphical analysis}

Before a formal analysis is performed, it is common and useful to informally inspect certain graphical plots for glaring issues with RD assumptions \citep{imbens2008regression,lee2010regression}. This inspection is briefly described here, with details and accompanying figures left to Appendix~\ref{sec:graphical}. First, RD design evaluates the effect of split flow by measuring the value of the discontinuity in the expected value of the outcome at the start of split flow. Plots of average outcomes are inspected to check that average outcomes are discontinuous at the start time of the split flow model \citep{imbens2008regression,lee2010regression}. Second, the continuity assumption can be incorrect if there is discontinuity in who arrives at $T=t_0$. Consequently, any discontinuity in observed outcomes could be attributed as much to this discontinuity in arrivals as to the patient flow model. Discontinuity in arrivals could arise if patients \emph{knew} when a split-flow model occurs and decide to manipulate their arrival time $T$ in order to be assigned a particular patient-flow model $A$ \citep{lee2010regression}. To examine this possibility, histograms of arrival times are inspected to look for discontinuity in number of arrivals before or after the start time of the split flow model. Further, if patients are unable to manipulate their arrival time, patient covariates (i.e., age, sex, and race) should, on average, be ``locally balanced'' on either side of start of split flow. Thus, plots of average covariates are inspected to look for discontinuities around the start time of the split flow model. To emphasize, the main concern is a \emph{discontinuity} in who arrives at $T=t_0$, not a continuous change over time such as when chest pain patients arrive earlier in the day than other types of patients.

\subsection{Bandwidth and polynomial order selection}

RD design involves a judgement for selecting the bandwidth $h$ and polynomial form of $f_0$ and $f_1$ \citep{lee2010regression}. On one hand, a large bandwidth could reduce variance of estimated $\gamma$ by increasing sample size. Polynomials of a small order might also reduce variance in estimated $\gamma$ as fewer coefficients are estimated. On the other, a small bandwidth and higher-order polynomials can improve the accuracy of the outcome model at Eq.~\eqref{eq:rho_model}. Moreover, a smaller bandwidth should yield ED visits that are more exchangeable and hence reduce estimation bias in average treatment effects. Thus, there is a clear trade-off between bias and variance when choosing the bandwidth. For reference, variation in ED visits has been studied on a scale of hours \citep{choudhury2020forecasting}, suggesting that the outcome model could be reasonably accurate with a bandwidth of about 1 hour, but may hide important ED trends with larger bandwidths. 

Following \cite{lee2010regression}, we used leave-one-out cross-validation in an effort to find the smallest bandwidth and suitable polynomial form that can yield stable predictions. For each outcome, this entailed fitting the model at Eq.~\ref{eq:rho_model} with given polynomial form of $f_0$ and $f_1$ to all ED visits within a given bandwidth $h$ of $t_0$ except for one visit within 1/2 hour of the split flow start time. Based on the fitted model, square error between the actual and predicted outcome was measured for the visit left out. This procedure was repeated leaving out each visit within 1/2 hour of the split flow start time. Square error was averaged over all observations within one-half hour of the split flow start time  and over all start times $t_0$ that are aggregated. 

Table~\ref{tb:rd_bandwidth} reports mean square error for bandwidth $h$ varied from 0.5 hours to 3 hours and polynomial form of $f_0$ and $f_1$ varied to be either linear with equal slopes, linear with different slopes, or quadratic. Reasonable mean square errors (when compared to other errors) could be achieved for all outcomes regardless of the polynomial choice $f_0$ and $f_1$ or the bandwidth. We thus focus on the smallest bandwidths of 1/2 hour and 1 hour with linear, equal-sloped polynomials. Other bandwidths will also be reported to check for sensitivity of findings to bandwidth selection.

\begin{table}[t!]
\centering
\begin{tabular}{c c c c c c c } 
\toprule
& \multicolumn{6}{c}{Bandwidth (hours)} \\
\cmidrule(lr){2-7}
Polynomial form & 0.5 & 1 & 1.5 & 2 & 2.5 & 3 \\
\midrule
& \multicolumn{6}{c}{Time to disposition} \\
\multirow{2}{*}{\makecell{  \textbf{Linear - same slopes} }} &  4.23 &  4.22 &  4.22 &  4.22 &  4.21 &  4.21 \\
&  (0.152) &  (0.153) &  (0.153) &  (0.154) &  (0.153) &  (0.153)     \\
\multirow{2}{*}{\makecell{  Linear - different slopes }} &  4.23 &  4.22 &  4.22 &  4.22 &  4.22 &  4.22   \\
&  (0.154) &  (0.153) &  (0.153) &  (0.159) &  (0.153) &  (0.153)     \\
\multirow{2}{*}{\makecell{  Quadratic }} &  4.23 &  4.22 & 4.22 & 4.22 &  4.22 &  4.22  \\
&  (0.153) &  (0.153) &  (0.154) &  (0.155) &  (0.154) &  (0.154)    \\
\midrule
& \multicolumn{6}{c}{ Triage time } \\
\multirow{2}{*}{\makecell{  \textbf{Linear - same slopes} }} &  0.19 &  0.19 &  0.19 &  0.20 &  0.20 &  0.21   \\
&  (0.011) &  (0.012) &  (0.012) &  (0.012) &  (0.012) &  (0.013)     \\
\multirow{2}{*}{\makecell{  Linear - different slopes }} &  0.19  & 0.19 &  0.19 &  0.19 &  0.19 &  0.19   \\
&  (0.011) &  (0.012) &  (0.012) &  (0.012) &  (0.012) &  (0.012)     \\
\multirow{2}{*}{\makecell{  Quadratic }} &  0.19 &  0.19 &  0.19 &  0.19 &  0.19 &  0.19   \\
&  (0.011) &  (0.011) &  (0.011) &  (0.011) &  (0.011) &  (0.012)     \\
\midrule
& \multicolumn{6}{c}{ Admission } \\
    \multirow{2}{*}{\makecell{  \textbf{Linear - same slopes} }} &  0.17 &  0.17 &  0.18 &  0.18 &  0.18 &  0.18   \\
&  (0.003) &  (0.003) &  (0.003) &  (0.003) &  (0.003) &  (0.003)     \\
\multirow{2}{*}{\makecell{  Linear - different slopes }} &  0.17 &  0.18 &  0.18 &  0.18 &  0.18 &  0.18   \\
&  (0.003) &  (0.003) &  (0.003) &  (0.003) &  (0.003) &  (0.003)    \\
\multirow{2}{*}{\makecell{  Quadratic }} &  0.17 &  0.17 &  0.17 &  0.17 &  0.17 &  0.17   \\
&  (0.004) &  (0.004) &  (0.004) &  (0.004) &  (0.004) &  (0.004)     \\
\midrule
& \multicolumn{6}{c}{ ED Revisit } \\
\multirow{2}{*}{\makecell{  \textbf{Linear - same slopes} }} &  0.11 &  0.10 &  0.10 &  0.10 &  0.10 &  0.10 \\
&  (0.004) &  (0.004) &  (0.004) &  (0.004) &  (0.004) &  (0.004) \\
\multirow{2}{*}{\makecell{  Linear - different slopes }} &  0.11 &  0.10 &  0.10 &  0.10 &  0.10 &  0.10\\
&  (0.004) &  (0.004) &  (0.004) &  (0.004) &  (0.004) &  (0.004) \\
\multirow{2}{*}{\makecell{  Quadratic }} &  0.12 &  0.10 &  0.11 &  0.10 &  0.10 &  0.10  \\
&  (0.004) &  (0.004) &  (0.004) &  (0.004) &  (0.004) &  (0.004) \\
\bottomrule
\end{tabular}
\caption{Mean square error (standard error) for four outcomes using leave-one-out cross-validation.}\label{tb:rd_bandwidth} \end{table}

\subsection{Addressing confounding, moderation, and mediation}
\subsubsection{Confounding variables.} Two approaches are used to account for the possibility that a shift and physician assignment confound the relationship between split flow and patient outcomes. We can attempt to indirectly determine the degree to which a shift change might impact patients outcomes by applying our RD analysis at start times of other shifts. For each of these times, we conducted the same RD analysis, but focused on patients within a bandwidth of these times and examined the effect of the shift change on patient outcomes. Meanwhile, to account for the possibility that a certain physician is a confounder, we repeated our RD analysis including a random intercept for each physician in our dataset and examined whether the effect of intervention significantly changed. The addition of a random intercept allows average outcomes to shift up or down according to which physician is assigned to patient, so that we can account for one physician who, e.g., yields faster treatment times than another. As before, we opted for a random intercept as opposed to a fixed intercept to reduce the number of parameters.

\subsubsection{Moderating variables.} We investigated, in three steps, the possibility of effect modification for each of the variables identified as a possible moderator in our causal framework (Section~\ref{sec:causal}). We first transformed congestion and physician workload into a categorical variable representing tertiles, with the lowest (highest) tertile corresponding to those days with the lowest (highest) levels of congestion or workload. The three tertiles for congestion correspond to ranges of $[2, 30]$, $(30, 41]$ and $(41, 81]$, whereas the three tertiles for physician workload correspond to ranges of $[0.2, 2.4]$, $(2.4, 3.1]$ and $(3.1, 17.7]$. For each moderator and each outcome, we next added indicators for each level of the moderator (e.g., low, middle, and high congestion tertile) and their interactions with the split flow indicator to the model described in Equation~\ref{eq:rho_model}. In each case, we used the fitted model to estimate average effects by level of a moderator and to perform a Wald hypothesis test of the null hypothesis that all interaction terms are zero. The resulting \textit{P}-value was used to indicate whether or not a variable was a significant moderator. Last, in an effort to recover the specific contributions of each moderator, we added all the indicators for every level of every moderator and their interactions with the split flow model to our RD analysis model. The final fitted model was used to estimate interaction terms by level of a moderator and to perform a Wald hypothesis test of the null hypothesis that all interaction terms for a given moderator are zero.

\subsubsection{Mediating variables.} For each mediator identified in Section~\ref{sec:causal}, we need to impose strong assumptions to identify average NIEs $\bar{\delta}(a)$ and NDEs $\bar{\zeta}(a)$. Following \cite{imai2010general}, we assume $\bar{\delta}(0)=\bar{\delta}(1)$ and $\bar{\zeta}(0)=\bar{\zeta}(1)$. We also assume the causal model in Figure~\ref{fig:DAG_Mediation} except there are no unmeasured variables confounding the relationship between the mediator $M$ and the outcome $Y$. That way, we can assume sequential ignorability \citep{imai2010general}: potential outcomes $Y^{a,b}$ and $M^{a'}$ are independent from $A$ conditioning on $X$ and potential outcomes $Y^{a',b}$ are independent from $M^{a'}$ conditioning on $X$ and $A$. Finally, our final assumption is a linear regression model for outcomes $Y$ and mediator $M$:
\begin{align}\label{eq:mediation_analysis}
\begin{split}
    \E[ Y | S, A, M, X] &= \alpha + \beta X + \psi S + \gamma A + \mu M \\
    \E[ M | S, A, X] &= \alpha' + \beta' X + \psi' S + \gamma' A.\\
\end{split}
\end{align}
This model for $Y$ is identical to original RD model at Equation~\ref{eq:rho_model} except with the addition of the mediator term and assuming a linear form with equal slope for $S$. A similar model is then used for the mediator $M$. Based on our assumptions, indirect effects are given by $\bar{\delta}(a)=\gamma'\mu$ and direct effects are given by $\bar{\zeta}(a)=\gamma$ \citep{imai2010general}. As before, we treat arrival day as a random effect. Additionally, we excluded encounters with an undefined mediator (i.e., 787  visits without a test order associated), the ignored encounters are primarily psychiatric visits, visits associated with suicidal behavior, and patients with eye problems. Last, we can also calculate mediating effects for different levels of a moderator by including the moderator and its interactions with the indicator of split flow into the models at Equation~\ref{eq:mediation_analysis} for $M$ and $Y$ as well as the interaction between the moderator and mediator for $Y$. 


\section{Results}\label{sec:results}

\smallskip

\subsection{Sample characteristics}

The analyzed sample (n = 21,570 visits) is summarized in Table~\ref{tb:sample} divided by whether the patient arrived to the ED in the 3 hours before (n = 9,965) or 3 hours after (n = 11,605) the start time of the split flow model. Briefly, patients were an average of 49.2 (SD = 19.2) years of age and were predominately white (79.3\%). Women visited the ED slightly more than men (54.7\%), and a majority of patients had either commercial insurance (44.1\%) or Medicare (30.1\%). The four outcomes of interest exhibited reasonable variability for analysis: time to be roomed has an average of 13.4 minutes (SD = 19.3); time to disposition has an average of 3.9 hours (SD = 2.0 hours); 24.5\% of visits were admitted to an inpatient hospital unit; and 11.9\% of visits led to an ED revisit within 30 days after being discharged from the ED or hospital.

\begin{table}[ht!]
\centering
\begin{tabular}{ l l r r r r}
\toprule
& & \multicolumn{2}{c}{\textbf{3 hours before split flow}}
& \multicolumn{2}{c}{\textbf{3 hours after split flow}} \\
& & \multicolumn{2}{c}{\textbf{n = 9,965}}
& \multicolumn{2}{c}{\textbf{n = 11,605}} \\
\cmidrule(lr){3-4} \cmidrule(lr){5-6}
\multicolumn{2}{c}{\textbf{Variable}} & \multicolumn{1}{c}{\textbf{Value}} & \multicolumn{1}{c}{\textbf{Missing}} &
\multicolumn{1}{c}{\textbf{Value}} & \multicolumn{1}{c}{\textbf{Missing}}\\
\midrule
\multicolumn{2}{l}{Age in years, mean (SD)} & 49.5 (19.2) & 0 & 48.9 (19.3) & 0 \\
\multicolumn{2}{l}{Time to be roomed in minutes, mean (SD)} & 12.7 (20.1) & 0 & 14.1 (18.7) & 0\\
\multicolumn{2}{l}{Time to disposition decision in hours, mean (SD)} & 3.9 (2.0) & 0 & 3.8 (2.1) & 0\\
\multicolumn{2}{l}{Female, n (\%)}  & 5,407 (54.2) & 0 & 6,404 (55.2)& 0\\ 
\multicolumn{2}{l}{Race and ethnicity, n (\%)}  &  & 0 &  & 0 \\ 
 & White & 7,904 (79.3) &  & 9,221 (79.4) & \\ 
 & Black & 1,044 (10.5) &  & 1,263 (10.9) & \\ 
 & Hispanic/Latino & 550 (5.5) &  & 573 (4.9) & \\  
 & Asian & 279 (2.8) &  & 331 (2.8) & \\  
 & Other & 80 (0.8) &  & 82 (0.7) & \\ 
 & American Indian/Alaska Native & 46 (0.5) &  & 60 (0.5) & \\  
 & Unknown & 62 (0.6) &  & 75 (0.7) & \\  
\multicolumn{2}{ l }{Health insurance, N (\%)} & & 0 & & 0\\ 
 & Commercial & 4,460 (44.7) &  & 5,050 (43.5) & \\  
 & Medicare & 2,980 (29.9) &  & 3,506 (30.2) & \\  
 & Medicaid/BadgerCare & 1,344 (13.5) &  & 1,713 (14.8) & \\  
 & Self paid & 425 (4.26) &  & 426 (3.67) & \\  
 & Unknown & 756 (7.6) &  & 910 (7.8) & \\  
\multicolumn{2}{l}{Chief complaint, n (\%)} &  &  0 & & 0\\
 & Abdominal pain & 1,183 (11.9) &  & 1,388 (11.9) & \\ 
 & Chest pain & 837 (8.4) &  & 823 (7.1) & \\ 
 & Dyspnea  & 525 (5.3) &  & 647 (5.6) & \\ 
 & Fall & 290 (2.9) &  & 320 (2.7) & \\ 
 & Fever & 171 (1.7) &  & 205 (1.8) & \\ 
 & Other & 6,959 (69.8) &  & 8,222 (70.8) & \\ 
 \multicolumn{2}{ l }{Admitted to hospital, n (\%)} & 2,385 (23.9) & 0 & 2,895 (24.9) & 0\\ 
 \multicolumn{2}{ l }{ED Revisit less than 30 days, n (\%)} & 1,200 (12.0) & 404 (4.0) & 1,372 (11.8) & 460 (3.9)\\ 
\bottomrule
\end{tabular}
\caption{
Sample characteristics of ED visits that met inclusion criteria in the 3 hours before and after the start time of the split flow model. } \label{tb:sample}
\end{table}

\subsection{Does split flow shorten length of stay}\label{sec:continuos_results}

Our primary RD analysis, which uses a 1 hour bandwidth, estimates that a split flow model significantly increases the average time to be roomed by $4.6$ minutes (95\% CI: [2.9, 6.2]) but significantly decreases average time to disposition decision after being roomed by an even larger margin of $14.4$ (95\% CI:[4.1, 24.7]) minutes (Table~\ref{tab:main_results}; these estimates are depicted in Appendix \ref{app:rd_figures}). Thus, the overall reduction in length of stay is 9.8 minutes. We reach similar conclusions when changing the bandwidth to 1/2 hour, though the magnitude of estimates are larger: a split flow model significantly increases the average time to be roomed by $9.1$ minutes (95\% CI: [6.7, 11.4]) but significantly decreases average time to disposition by $20.4$ (95\% CI:[6.1, 34.7]) minutes. The overall reduction in length of stay, however, is similar at 11.3 minutes. Estimates for other bandwidths are reported in Appendix~\ref{app:bandwidths}. Taken together, these findings suggest that a split flow model does indeed shorten length of stay, as was hypothesized.

\begin{table}[ht!]
    \centering
    \begin{tabular}{l c c}
    \toprule
            \textbf{Analysis} & \textbf{Time to be roomed} & \textbf{Time to disposition} \\
        \midrule
        RD with 1 hour bandwidth (primary) & 4.6 (2.9, 6.2) & -14.4 (-24.7, -4.1) \\
        & \multicolumn{2}{c}{\ul{Sensitivity analyses}} \\
        Changing bandwidth to 1/2 hour & 9.1 (6.7, 11.4) & -20.4 (-34.7, -6.1)\\
        Controlling for physician & 4.5 (2.8, 6.1) & -14.7 (-25.0, -4.3) \\
    \bottomrule
    \end{tabular}
    \caption{Estimates (95\% confidence intervals) for average effects of the split flow model on time to be roomed and time to disposition decision after being roomed, in minutes.}
    \label{tab:main_results}
\end{table}

We performed additional analyses to investigate the the possibility of confounding. As we noted earlier, one possible confounder was the physician assigned to split flow, since certain physicians may be faster than others. To account for variation in physicians, we added assigned physician as a random effect to model~\ref{eq:rho_model} (Table~\ref{tab:main_results}). Compared to our primary estimates, accounting for physician variation leads to an estimated average time to be roomed that increases by a slightly smaller amount of $4.5$ minutes (95\% CI: [2.8, 6.1]) and an estimated average time to disposition decreased by a slightly bigger amount of $14.7$ minutes (95\% CI: [4.3, 25.0]). These results suggest that reduction in length of stay under a split flow model is not due to the physician assigned to split flow.

Another possible confounder is a change in shift of physician providers in the ED, since split flow coincides with a shift change. However, when we estimate the same model by replacing $t_0$ with the start/end hour of other shift changes, we do not find a significant effect of a shift change on time to be roomed and time to disposition (Table~\ref{tab:tt_shift_changes}). That is, confidence intervals contain zero for each start/end hour of each shift.  These findings suggest that the reduction in LOS under a split flow model is not due to a shift change.

\begin{table}[ht!]
    \centering
    \begin{tabular}{l c c c c c c}
    \toprule
         & \multicolumn{2}{c}{\textbf{Shift 1}} & \multicolumn{2}{c}{\textbf{Shift 2}} & \multicolumn{2}{c}{\textbf{Shift 3}}\\
         \cmidrule(lr){2-3} \cmidrule(lr){4-5} \cmidrule(lr){6-7}
         \textbf{Outcome} & 7am & 4pm & 3pm & 12pm & 11pm & 8am\\
         \midrule
         Time to be roomed & -0.03 & -0.7 & 0.7 & 0.0 & -1.7 & -0.3\\
         & (-0.8, 0.8) & (-1.6, 0.1) & (-0.1, 1.5) & (-2.9, 2.8) & (-4.6, 1.1) & (-0.1, 0.3)\\
         Time to disposition & -4.0  & 9.7  & -5.3  & 3.3 &  -2.9 & 9.6 \\
         & (-23.3, 15.2) & (-1.4, 20.8) & (-16.1, 5.5) & (-11.8, 18.5) & (-17.0, 11.2) & (-7.1, 26.4)\\
         \bottomrule
    \end{tabular}
    \caption{Estimates (95\% confidence intervals) for average effects of the split flow model on outcomes when the forcing variable is centered at the start/end of shift changes, in minutes.}
    \label{tab:tt_shift_changes}
\end{table}

\subsection{When does split flow shorten length of stay (moderation analyses)}
We next investigated whether the split flow was more or less effective depending on the operational context of the ED (i.e., congestion, day of the week, physician workload, and start time of the split flow model). When adding, one at a time, each possible moderator and their interaction with the split flow indicator to the model, we found that the start time of the split flow model did not yield significant interaction terms in the model of time to be roomed (\textit{P} = 0.08) or in the model of time to disposition (\textit{P} = 0.48). Therefore, the split flow start time is not a significant moderator for either time to be roomed or time to disposition decision after being roomed. Similarly, day of week (\textit{P} = 0.06) nor physician workload (\textit{P} = 0.07) yielded significant interaction terms in the model of time to disposition. Therefore, day of week nor physician workload appear to moderate the effect of split flow on time to disposition.

\begin{figure}[ht!]
\centering
\includegraphics[width=\textwidth]{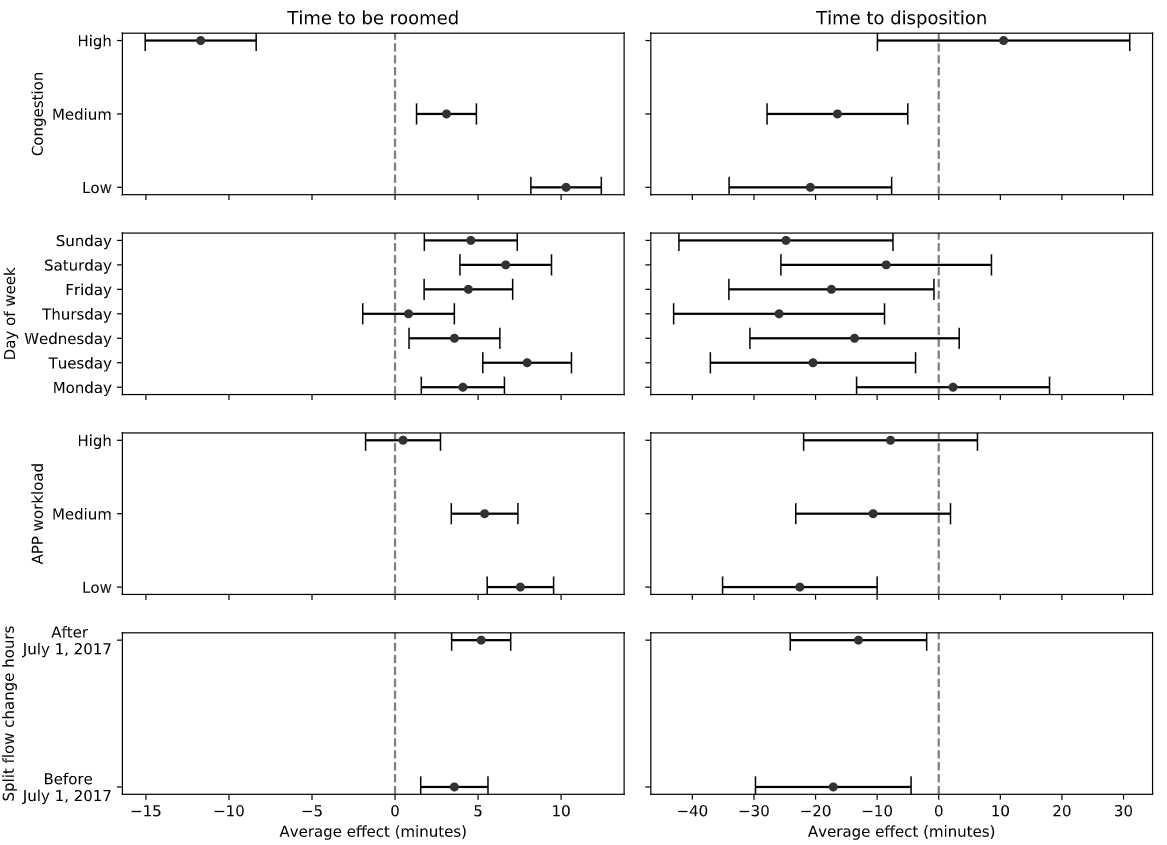}
\caption[]{Moderation analyses for continuous outcomes. Average effect (in minutes) of split flow on time to be roomed (left) and time to disposition after being roomed (right) during each level of moderator variables: congestion, day of week, physician workload and split flow change of start hours.}
\label{fig:moderation_continuous_outcomes}
\end{figure}

By contrast, congestion (\textit{P} $<$ 0.001), day of week (\textit{P} = 0.001) and physician workload (\textit{P} $<$ 0.001) did yield significant interactions for time to be roomed. The only significant moderator for time to disposition was congestion (\textit{P} = 0.01). Thus, the effectiveness of the split flow model does appear to be moderated by congestion, day of week, and physician workload. Upon closer examination, we find that the split flow model increases average time to be roomed during the low congestion level, and to a lesser degree, during the middle congestion level (Figure~\ref{fig:moderation_continuous_outcomes}). However, the split flow model actually decreases average time to be roomed during the high congestion level.  This alone might suggest that split flow is more effective during high congestion levels, but time to disposition tells a different story. In this case, we find that the split flow model decreases average time to disposition during the low congestion and middle congestion levels. Yet when we take both time to be roomed and time to disposition after being roomed into consideration, these findings suggest that split flow is especially effective at low and medium congestion levels.

We find that physician workload moderates the relationship between split flow and time to be roomed/disposition in a similar way as congestion (Figure~\ref{fig:moderation_continuous_outcomes}). We find that split flow increases the time to be roomed but to a greater degree decreases the time to disposition during low physician workload levels. These findings suggest that split flow decreases LOS during low congestion levels. On the other hand, we find that split flow increases time to be roomed while not impacting time to disposition when physician workload is at medium levels. These results taken together suggest that split flow increases ED LOS during medium physician workload levels. We find no significant evidence of split flow either increasing or decreasing time to be roomed or time to disposition for the highest tertile of physician workload.

As for day of the week, the average effects of split flow on time to be roomed and on time to disposition are similar across days of the week (note the overlapping confidence intervals in Figure~\ref{fig:moderation_continuous_outcomes}), with a few exceptions. Under a split flow model, average time to be roomed increases more on Tuesday and less on Thursday than other days, whereas average time to disposition decreases less on Monday. Overall, split flow appears to be most effective on Thursday and least effective on Monday. We suspect that this variation is largely due to differences in how busy the ED is on different days, which we explore next.

Since these moderators may relate to each other, we wanted to see which might be driving the moderation of average effects. We therefore simultaneously added congestion, day of the week, physician workload, and split flow start hours and their pairwise interactions with the split flow indicator to the model. From this final model, we again find that split flow start hours is not a significant moderator (\textit{P} = 0.08 for time to be roomed and \textit{P} = 0.40 for time to disposition).  Physician workload, which was a significant moderator when considered in isolation, did not yield significant interaction terms with either time to be roomed (\textit{P} = 0.17, Table~\ref{tb:moderation_full_model}) or time to disposition (\textit{P} = 0.85). Unlike physician workload, congestion (\textit{P} $<$ 0.001) and the day of week (\textit{P} = 0.007) remain an important moderator of time to be roomed yielding highly significant interaction terms. However, neither congestion (\textit{P} = 0.06) nor day of week (\textit{P} = 0.14) yielded significant interactions with time to disposition. These findings suggests that regardless of multiple moderator variables being significant individually, the most important moderator of LOS is congestion followed by day of week. Specifically, the time to be roomed is moderated by congestion and day of week. Split flow, for example, increases the time to be roomed on Tuesdays (the confidence intervals for all days contain zero except Tuesday). Most notably, split flow is most effective during low and medium congestion levels, which is contrary to our original hypothesis that split flow would be most effective during high congestion levels.

\begin{table}[ht]
    \centering
    \begin{tabular}{l l c c c c}
        \toprule
        & & \multicolumn{2}{c}{\textbf{Time to be roomed}} & \multicolumn{2}{c}{\textbf{Time to disposition}}\\
        \cmidrule(lr){3-4} \cmidrule(lr){5-6}
         \textbf{Moderator} &  Level & Interaction (95\% CI) & $P$ & Interaction (95\% CI) & $P$\\
        \midrule
         Congestion & Medium & -8.3 (-10.8, -5.9) &  \multirow{2}{*}{\makecell{ $<$ 0.001 }} & 1.7 (-13.5, 16.8) & \multirow{2}{*}{\makecell{ 0.06 }}\\
         & High & -23.9 (-28.0, -19.8) & & 27.1 (1.4, 52.7)\\
         \cmidrule(lr){2-6}
         Day of week & Monday & 2.0 (-1.2, 5.1) & \multirow{6}{*}{\makecell{ 0.007 }} & 18.9 (-1.5, 39.3) & \multirow{6}{*}{\makecell{ 0.14 }}\\
         & Tuesday & 4.1 (0.8, 7.4) & & -1.2 (-22.2, 19.7)\\
         & Wednesday & -0.2 (-3.5, 3.0) & & 7.8 (-13.2, 28.8)\\
         & Thursday & -2.2 (-5.5, 1.1) & & -5.4 (-26.7, 16.0)\\
         & Friday & 0.3  (-2.9, 3.5) & & 2.0 (-18.7, 22.7)\\
         & Saturday & 0.8 (-2.4, 4.1) & & 16.0 (-4.7, 36.6)\\
         \cmidrule(lr){2-6}
         Physician workload & Medium & -6.1 [-8.3, -4.0] & \multirow{2}{*}{\makecell{ 0.17 }} & 4.0 (-10.3, 18.3) & \multirow{2}{*}{\makecell{ 0.85 }}\\
         & High & -5.8 (-9.8, -1.9) & & 1.8  (-17.1, 20.7) \\
         \cmidrule(lr){2-6}
         Split flow start hour & After July 1, 2017 & -1.7 (-3.5, 0.2) & 0.08 & 5.0 (-6.7, 16.7) & 0.40\\
         \bottomrule
    \end{tabular}
    \caption{Estimates for the coefficients of the interaction terms between split flow and moderator variables for the regression model including all moderators, and $P$ values for the null hypothesis testing whether the interactions of a given moderator are zero simultaneously.}
    \label{tb:moderation_full_model}
\end{table}

\subsection{Does split flow have downstream consequences}\label{sec:binary_results}
The previous section estimated an overall decrease in average length of stay.  This section explores whether this improvement in operational outcomes is at the cost of downstream consequences.  In particular, we evaluate the impact of split flow on admission and revisit rates. The results in Table~\ref{tab:secondary_results} suggest that a split flow model significantly reduces the probability of admission by $5.8\%$ (95\% CI: [2.3\%, 9.4\%]) but does not have a significant effect on 30-day revisit rates. Changing the bandwidth to 1/2 hour yields similar conclusions, i.e. split flow leads to a significant decrease in admission rates but no significant change in 30-day revisit rates (Table~\ref{tab:secondary_results}). Like we did for time to be roomed and time to disposition, we also explored whether the estimates are confounded by assigned physician or shift changes. For the former, we obtained similar results after adding the physician as a random effect (Table~\ref{tab:secondary_results}). For example, a split flow model reduces the admission rate by $5.6\%$ (95\% CI: [2.0\%, 9.2\%]), whereas the effect on the probability of revisit remains not significant. Similarly, the confidence intervals for each RD estimate when the forcing variable is centered at the start/end of shifts changes are not significant (i.e., contain zero) (Table~\ref{tab:binary_shift_changes}), which suggests that a shift change is not sufficient to explain that impact that split flow has on admission rates. As before, estimates for other bandwidths are reported in Appendix~\ref{app:bandwidths}, discontinuities in estimates are depicted in Appendix~\ref{app:rd_figures}, and the moderation analyses for downstream outcomes are reported in Appendix~\ref{app:moderation_binary}.

\begin{table}[ht!]
    \centering
    \begin{tabular}{l c c}
    \toprule
            \textbf{Analysis} & \textbf{Admission decision} & \textbf{30-day revisit} \\
        \midrule
        RD with 1 hour bandwidth (primary) & -5.8 (-9.4, -2.3) & -0.8 (-3.6, 1.9) \\
        & \multicolumn{2}{c}{\ul{Sensitivity analyses}} \\
        Changing bandwidth to 1/2 hour & -7.9 (-12.8, -3.0) & -1.6 (-5.4, 2.2)\\
        Controlling for physician & -5.6 (-9.2, -2.0) & -0.7 (-3.5, 2.0) \\
    \bottomrule
    \end{tabular}
    \caption{Estimates (95\% confidence intervals) for average effects of the split flow model on admission decision and 30-day revisits, in \%.}
    \label{tab:secondary_results}
\end{table}



\begin{table}[ht!]
    \centering
    \begin{tabular}{l c c c c c c}
    \toprule
         & \multicolumn{2}{c}{\textbf{Shift 1}} & \multicolumn{2}{c}{\textbf{Shift 2}} & \multicolumn{2}{c}{\textbf{Shift 3}}\\
         \cmidrule(lr){2-3} \cmidrule(lr){4-5} \cmidrule(lr){6-7}
         \textbf{Outcome} & 7am & 4pm & 3pm & 12pm & 11pm & 8am\\
         \midrule
         Admission decision & -1.7 & 0.9 & -1.7 & 3.4 & -2.0 & 3.4\\
         & (-7.8, 4.4) & (-4.8, 2.9) & (-5.5, 2.1) & (-1.5, 8.4) & (-6.3, 2.4) & (-2.0, 8.7)\\
         30 day revisit & -1.0  & -0.6  & -0.4 & 5.3 & -3.5 & -0.7 \\
         & (-6.6, 4.6) & (-3.8, 2.6) & (-3.5, 2.7) & (0.6, 10.1) & (-7.6, 0.6) & (-5.4, 3.9) \\
         \bottomrule
    \end{tabular}
    \caption{Estimates (95\% confidence intervals) for average effects of the split flow model on admissions and 30 day revisits when the forcing variable is centered at the start/end of shift changes, in \%.}
    \label{tab:binary_shift_changes}
\end{table}

\subsection{Why does split flow shorten length of stay (mediation analysis)}

Given that split flow may be reducing time to disposition, we now explore why that might be, especially as it relates to treating low-needs patients in a vertical area vs. triage and early task initiation by a physician. We analyzed two variables, time to first order of a test (e.g., electrocardiogram, radiology) or medication and time to be roomed, that might mediate the relationship between split flow and time to disposition. We find that a decrease of 2.0 minutes (95\% CI: [0.3, 4.1]) in time to disposition from split flow can be attributed to quicker test and medication orders under split flow compared with the traditional model (Table~\ref{tab:results_mediation}). When broken down by congestion level, this indirect effect of split flow via time to first order is stronger at higher congestion levels, e.g. 2.8 minutes for the high congestion tertile. In other words, providers are able to initiate medication and test orders earlier under split flow which decreases time to disposition, especially in periods of high congestion. So, early task initiation by a physician appears to partly explain why split flow is effective.

The other mediator, time to be roomed, brings less clarity about why split flow is effective. Our estimates suggest that a slight decrease of 0.5 minutes (95\% CI: [0.002, 0.8] minutes) in time to disposition can be attributed to slower time to be roomed under split flow compared to traditional flow (Table~\ref{tab:results_mediation}). One possible explanation for why a slower time to be roomed under split flow might contribute to faster time to disposition is that the physician spends time under split flow to initiate treatment (including medication and test orders) before patients are officially roomed. While perhaps plausible at medium congestion levels, this explanation seems to fall apart at low and high congestion levels. At low congestion levels, slower time to be roomed under split flow causes a slower --- not faster --- time to disposition, resulting in a non-significant increase of 1.5 minutes (95\% CI: [-2.6, 7.3] minutes) in time to disposition attributed to slower time to be roomed under split flow. Meanwhile at high congestion levels, split flow causes a faster --- not slower --- time to be roomed, suggesting the physician assigned to triage is not spending extra time to initiate care.

An alternative explanation is that the physician assigned to triage adjusts how they stream patients according to congestion level, which in turn changes how patients are handled by physicians downstream in the ED. At high congestion, such an explanation would mean that patients are more quickly streamed to rooms (or the vertical area) by the physician at triage under a split flow model but are not treated at a faster rate, causing a slower time to disposition and a significant increase of 3.2 minutes (95\% CI: [0.7, 7.3] minutes) in time to disposition from the faster time to be roomed under split flow. At low congestion, this explanation would mean that patients are streamed more slowly to a room by the physician at triage than a nurse at triage and that patients who are moved to a room more slowly are also slower to receive a disposition decision. Thus, these findings suggest that the physician at triage engages in patient streaming to modulate the rate at which patients are roomed and treated, but it remains unclear if patient streaming, unlike early test or medication orders, helps to speed up patient flow in the ED.

\begin{table}[ht!]
    \centering
    \begin{tabular}{l l c c c c}
        \toprule
        \textbf{Mediator} & \textbf{Group}  & \textbf{Direct effect} & \textbf{Indirect effect} \\
        \midrule
         \multirow{4}{*}{\makecell{ Time to first order }}  & All & -11.9 (-23.9, -2.0)  & -2.0 (-4.1, -0.3)\\
         & Low congestion & -19.0 (-32.1, -7.3) & -1.6 (-5.5, 0.4)\\
         & Medium congestion& -13.4 (-24.4, -13.5) & -2.4 (-4.1, -0.6) \\
         & High congestion & 15.8 (-1.12, 39.24) & -2.8 (-7.0, 1.6) \\
         \cmidrule(lr){2-4}
         \multirow{4}{*}{\makecell{ Time to be roomed }}  & All & -13.3 (-14.8, -3.6) & -0.5 (-0.8, -0.002)\\
         & Low congestion& -19.0 (-33.1, -8.0) & 1.5 (-2.6, 7.3)\\
         & Medium congestion& -15.4 (-29.1, -3.4) & -0.5 (-1.2, -0.02)\\
         & High congestion& 8.4 (-9.1, 27.5) & 3.2 (0.7, 7.3) \\
         \bottomrule
    \end{tabular}
    \caption{Estimates (95\% confidence intervals) for the average direct effect and average indirect effect of the split flow model on time to disposition mediated the time to first order and time to be roomed per congestion level, in minutes.}
    \label{tab:results_mediation}
\end{table}

\section{Discussion}\label{sec:conclusion}

In this paper, we evaluated a split flow model at an ED in a large tertiary academic hospital using electronic health records (n = 21,570). Split flow models, in which a provider rather than a nurse handles triage, have been proliferating in recent years as a possible strategy for improving ED patient flow and other operational metrics without sacrificing patient safety.  The present work is based on the premise that to determine whether or not such interventions should be continued, adapted, or more widely adopted by hospitals, we should understand their impact on operational metrics and patient outcomes from places where split flow models have been implemented. We hypothesized that a split flow model would lead to an improvement on patient flow, but no impact on downstream patients outcomes. 

We estimated average treatment effects of a split flow vs. traditional nurse-led triage flow model of the ED on time to be roomed, time to disposition decision after being roomed, the admission decision, and revisit to the ED within 30 days of discharge. These outcomes measure patient flow, decision-making, and subsequent patients outcomes. Estimates were recovered from a regression discontinuity (RD) design. We find an small increase in time to be roomed but a greater reduction in time to disposition. Taken together, they suggest a moderate reduction in average length of stays per patient. RD design, for example, found a split flow model leads to an estimated increases in average time to be roomed of 4.6 minutes (95\% CI: [2.9, 6.2] minutes) and a reduction in average time to disposition of 14.4 minutes (95\% CI: [4.1, 24.7] minutes). 
Although our findings support our hypothesis that a split flow model improves patient flow, we had also hypothesized that a split flow model would have not have downstream consequences to admission rates or ED revisits. As hypothesized, we find no evidence of a significant impact of split flow on revisit rates. However, our findings suggest a split flow model may decrease ED admissions by 5.9\% (95\% CI: [2.3\%, 9.5\%]). Since admitting a patient is one of the most costly decisions made in healthcare, then fewer admission rates may be viewed as an unexpected benefit of split flow.  In short, split flow model appears to reduce average lengths of stay and admission rates, without causing an increase in ED revisits.

For hospital mangers who might consider implementing a split flow model, it is important to determine when a split flow model might be especially effective. Congestion appears to be the most important factor modifying effectiveness. We conjectured that split flow would be less effective during periods of high congestion, considering that physicians already speed up care during periods of high congestion. Indeed, we found that split flow is most effective at reducing length of stay during periods of low congestion and relatively ineffective during periods of high congestion.

When congestion is low, patients spend more time waiting for a room under split flow vs. traditional flow, which provides the physician assigned to triage an opportunity to initiate treatment for patients. Once roomed, physicians could reach a disposition decision more quickly under split flow, perhaps because of the early initiation of care. The speed-up in time to disposition was greater than the delay in being roomed, leading to an overall reduction in length of stay. When congestion is high, however, patients spend less time waiting for a room under split flow vs. traditional flow. Here, we suspect the physician in triage is able to quickly stream low-needs patients to the vertical area, freeing up rooms for other patients. However once roomed, the time to disposition is slower under a split flow model, perhaps because patients are still competing for resources even in the vertical area. In the end, the gain in time to be roomed is almost equal to the loss in time to disposition when congestion is high, leading to a negligible net gain. Regardless, these results point to the use of split flow to improve patient flow during periods of low and medium congestion.

We acquired further empirical support for these insights using mediation analysis. This analysis suggested that the impact of split flow on time to disposition is partly mediated by the ability of physicians to order tests and medication more quickly when compared with the traditional nurse-led triage. Put simply, split flow led to early task initiation, which allowed a disposition decision to be reached sooner. Mediation analysis also suggested that, at high congestion levels, the impact of split flow time on time to disposition is partly mediated by time to be roomed. In this case, split flow led to faster time to be roomed, which led to slower time to disposition. So while the physician at triage appears to move patients more rapidly to the vertical area at high congestion levels, it is not clear if treating patients in this area yields net gain in length of stay.

Considering that estimates are sensitive to modeling assumptions, it is important to search for evidence of violations of our assumptions in an effort to strengthen the validity of these estimates. One concern is that split flow coincides with other operational changes, notably a change in the physician shift and the assignment of a physician, which might confound the relationship between split flow itself and patient outcomes. However, accounting for physician assignment had a minimal impact on our estimates. Further, shift changes for all physicians, other than split flow, did not seem to affect patient outcomes significantly. Thus, we believe that the effects can be largely attributed to the split flow model itself. Another concern is our key assumption about continuous potential outcomes. To that end, we investigated bandwidth selection, which we found did change the magnitude of our estimates but not our conclusions. We also performed graphical analysis, but did not observe a considerable amount of patients accumulating on one side of the cutoff. This fact favors the assumption of continuous potential outcomes. The graphical analysis on the covariates included in the regression model, namely, age, race, sex and chief complaint suggest that they are all continuous at the start of the split flow model except for a possible discontinuity in the average age of patients arriving before and after the cutoff. So, we would caution the reader when interpreting estimates as one does all attempted causal inferences made from observational data rather than from a randomized control trial, before making recommendations for continuing, adapting, and/or adopting split flow models.  

For example, even if a split flow model leads to an overall improvement in patient flow, a hospital administrator may wish to consider other factors when deciding whether to implement a split flow model. One consideration is patient preference. More specifically, managers might want to implement split flow when congestion is high, even if split flow does not lead to an overall reduction in length of stay, since patients may strongly prefer waiting in an ED treatment room over waiting in the ED waiting room. Another consideration is whether a decrease in average length of stay of about 8-10 minutes is sufficient in size to justify a change to how triage is implemented. We note that 8-10 minutes per patient may seem like a small reduction, but when accumulated over how many patients are seen over several hours, it may provide meaningful time savings to providers and hospital managers.  For example, our partner hospital roughly sees 163 patients per day on average. A benefit of the present study is that the provided estimates could be used as input for a cost-benefit analysis of if and when to adapt, continue, or adopt a split flow model.  In particular, hospital managers and care providers need to weigh the benefit of reducing average ED length of stay by some time unit against the potential additional cost of hiring an additional provider.

To our knowledge, we are the first apply causal inference methods, notably an RD analysis, to evaluate average treatment effects of split flow model on patient flow and patient outcomes. Prior empirical support for split flow models have focused on using more traditional regression methods to analyze the impact of split flow models on decreased length of stays \citep{subash2004team,medeiros2008improving,han2010effect,Soremkun12,Burstrom12,burstrom2016improved}. Meanwhile, empirical support for sub-components of split flow models, i.e., physician-triage and treating low-needs patients in a vertical area, have focused on matching strategies to adjust for measured confounding \citep{russ2010placing,partovi2001faculty,traub2016physician,considine2008effect}. Because patients are not randomized to the patient flow model, traditional regression and matching methods can be biased by unmeasured differences between intervention groups (split flow vs. traditional flow). RD analysis benefits from its potential to adjust for both measured and unmeasured confounding. We thus believe findings from a RD analysis should serve to strengthen the support for implementing a split flow model.


In addition, the intervention of interest (i.e. split flow model) had the interesting feature of being implemented over a specific period of time each day of the week. This feature does not fit within a sharp RD design, which has one threshold for determining whether or intervention condition (as opposed to a threshold for each day). To place our intervention into a RD framework, we centered each arrival at the start of split flow and aggregated all the visits in our dataset across days. 
 
There are several limitations to consider. First, we cannot test the continuity of potential outcomes purely from observational data. To alleviate this, we performed  graphical analysis to search for evidence of a discontinuous distribution of arrivals near the cutoff that could invalidate our results. Second, we cannot generalize our findings to hours outside the 1 hour bandwidth around the start of split flow. The split flow may be effective in reducing average length of stay just at the start but in fact increase time to disposition several hours after its implementation. Indeed, we report an RD analysis in the Appendix applied to the end time of split flow and find that split flow is less effective in reducing length of stay at the end hour.

In addition, the aggregation of observations across days helps increase the RD sample size but at cost of, possibly, hiding seasonal factors that influence patient arrival processes. Another limitation is that outside environmental factors may have influenced the effectiveness of the split flow. These include a time-lag until split flow protocols were standardized after its initial implementation. Further, certain choices made in data analysis may be considered limitations. Confidence intervals may have understated uncertainty in estimates. We defined ED revisits based on time of discharge from hospital for admitted patients and time of ED discharge for discharged patients, but alternatively the latter time could be used for both. Last, our sample is relatively homogeneous, limiting the generalizability of our findings to more diverse populations.   

In summary, our work extends our understanding of the benefits from ED split flow models. Finding overall reductions in ED length of stay without negatively impacting admission decisions and revisits may support the adoption (or adaption) of split flow models.  Our results do not support a change in current clinical practice, especially in light of the considerations mentioned above; rather they add to a growing body of literature suggesting that split flow models may yield benefits, and point to the necessity of further assessment to identify specific criteria to guide the adoption of such models. Lastly, our analysis is made possible with an approach which can be used to assess other interventions that are administered at specific times each day.

\bibliographystyle{unsrtnat}
\bibliography{references}  






\begin{appendices}

\section{Graphical Analysis} \label{sec:graphical}

\begin{figure}[ht!]
\centering
\includegraphics[width=\textwidth]{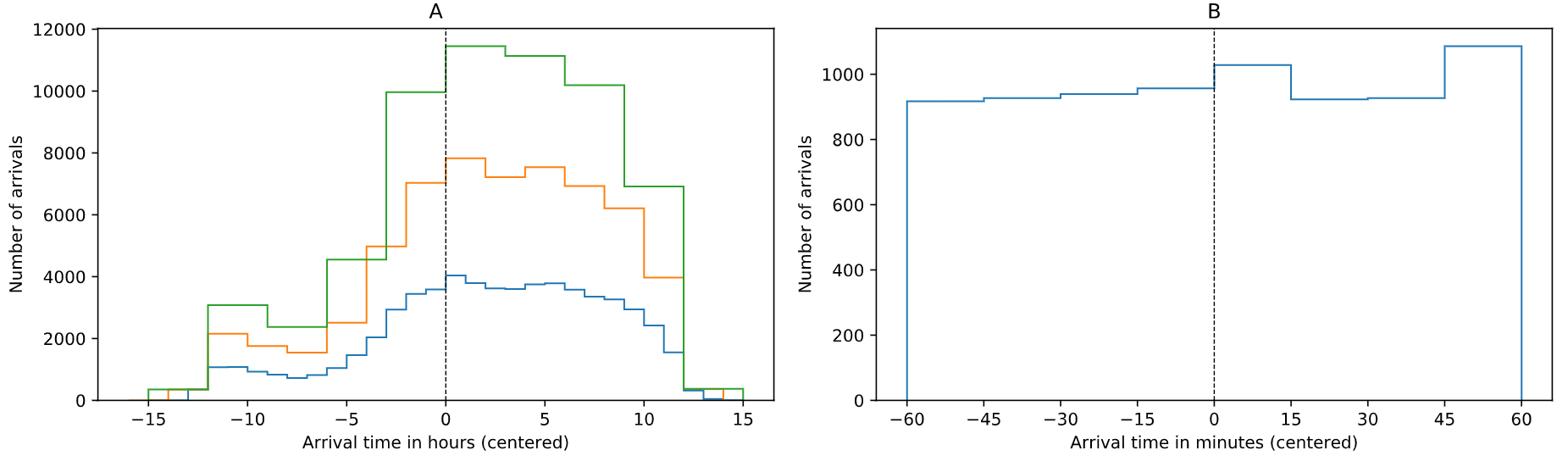}
\caption[]{Histograms of the number of arrivals for all the sample (Panel A) and the 1 hour bandwidth used in the RD analysis (Panel B). Non-overlapping bins of 1 (blue), 2 (orange) and 3 (green) hours were used in Panel A and 15 minutes bins were used for Panel B. The dashed vertical line marks the start of the split flow model.}
\label{fig:forcing_variable_hist}
\end{figure}

Following \cite{imbens2008regression} and \cite{lee2010regression} for RD designs, we present several plots to look for glaring violations of the Continuity Assumption~\ref{ass:rd2} and to confirm the possibility of a non-zero average effect of the split flow model on outcomes. Figure \ref{fig:forcing_variable_hist} is a histogram of arrival times for the entire sample (Panel A) and the 1 hour bandwidth (Panel B) used in the RD analysis. For the figure in Panel A, non-overlapping bins around the cutoff of sizes of 1 (blue), 2 (orange) and 3 (green) hours were used. For the figure in Panel B, non-overlapping bins of 15 minutes were used. These histograms show jumps near the start of split flow that are not substantially larger than any other time of day. We also do not see any notable accumulation of patients before and after the start of split flow. For example, the number of patients one hours before and one after the split flow model start time was 3590 and 4215 respectively across the RD sample. These findings do not suggest that patients are manipulating their arrival times to receive a certain patient flow model. If indeed patients are not manipulating arrival time, then it is also reasonable to assume potential outcomes are continuous at the start time of the split flow model, as assumed in Assumption~\ref{ass:rd2}.

\begin{figure}[ht!]
\centering
\includegraphics[width=\textwidth]{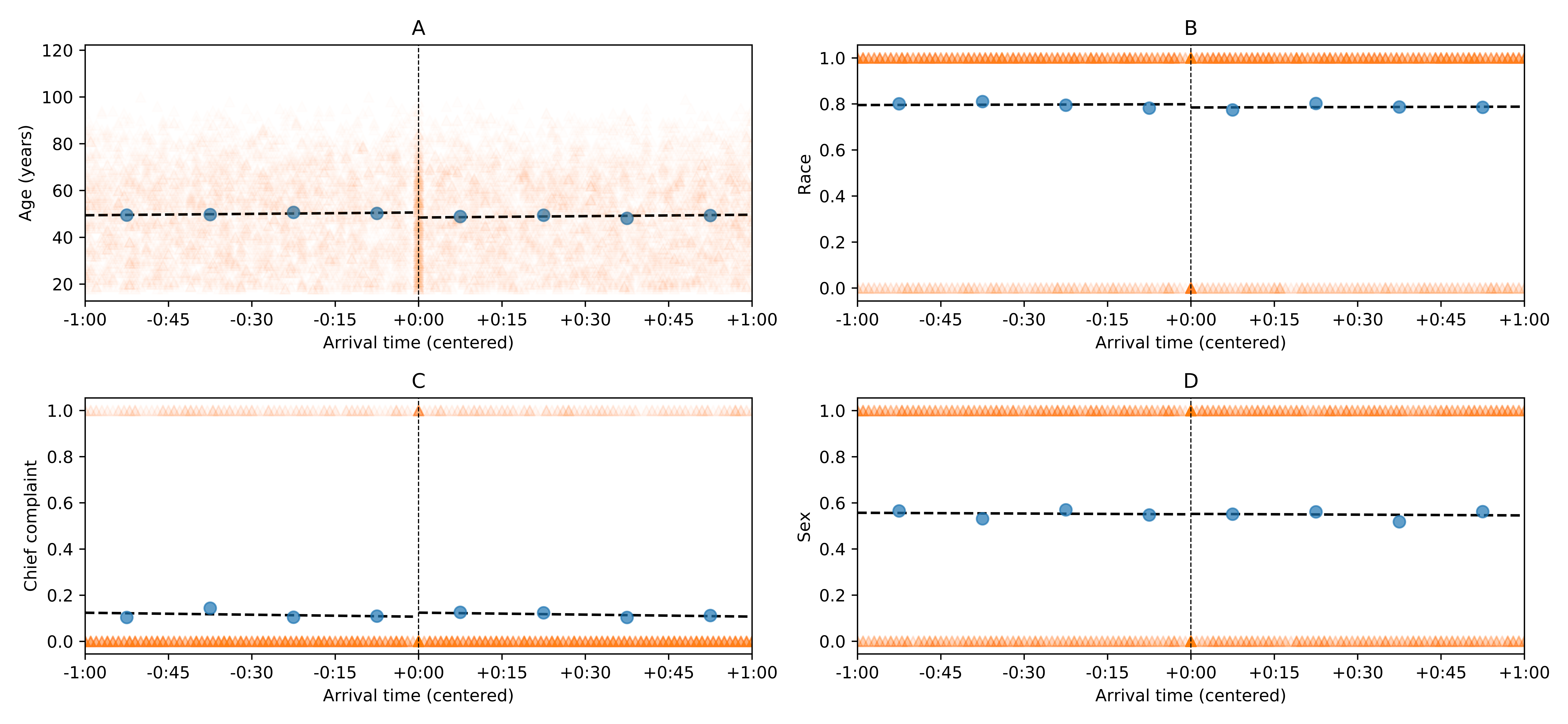}
\caption[]{Covariate balancing. Panels A, B, C and D represent the average age, race, chief complaint and sex (blue dots) for different visits (orange triangles) in the 1 hour bandwidth sample. Pre-treatment variables were averaged in bins of 15 minutes. The dashed vertical line marks the start of the split flow model.}
\label{fig:time_covariates}
\end{figure}

As an additional check for treatment manipulation, Figure \ref{fig:time_covariates} plots mean values of age, race, chief complaint and female indicator (Panels A, B, C and D respectively) for the RD sample. Each figure consists of non-overlapping bins of size 15 minutes centered at the start of split flow (dashed vertical line). Importantly, the start of the split flow model is used as one of the boundaries between two consecutive bins in order to prevent overlap of split flow patients with traditional flow patients. The presence of a jump at the cutoff could invalidate the RD design since they affect the estimated discontinuity for the outcome. These plots shows jumps in average covariates near the start of split flow that are not substantially larger than any other time of day except for age, where we find that average age decreases by 1.9 years (95\% CI:[0.3, 3.5], \textit{P} = 0.017) between control and treatment groups (see Table~\ref{tab:rd_covariates}). This alone suggests that age might be discontinuous at the start of split flow. However, when we account for multiple comparisons, as recommended by \cite{lee2010regression}, age is not significant with a Bonferroni-corrected \textit{P} value of 0.017 $\times$ 4 = 0.068. Thus, it does not appear that patients with certain characteristics are manipulating their arrival times, which is consistent with Assumption~\ref{ass:rd2}. As further support, we note that adding and removing covariates barely changed the estimations (see Table~\ref{tab:multiple_bandwidth}).

\begin{table}[ht!]
    \centering
    \begin{tabular}{l c c c}
    \toprule
        Covariate & RD estimate & 95\% Confidence interval & $P$\\
        \midrule
        Age (years) & -1.9 & [-3.5, -0.3] & 0.017\\
        Race & -0.5\% & [-3.9\%, 2.9\%] & 0.78\\
        Chief complaint & 1.5\% & [-1.2\%, 4.2\%] & 0.94\\
        Sex & 0.2\% & [-4.4\%, 4.1\%] & 0.28\\
    \bottomrule
    \end{tabular}
    \caption{Estimates (95\% confidence intervals) of a discontinuity when replacing the outcome for each covariate of the base model. Here, $P$ values are provided before applying a Bonferroni correction to account for multiple comparisons.}
    \label{tab:rd_covariates}
\end{table}

\newpage

\section{Sensitivity to bandwidth and covariates}\label{app:bandwidths}

Bandwidth selection is critical to RD designs. The bandwidth size is intimately related to the trade-off between bias and variance associated with the RD estimates. A narrow bandwidth increases variance, but allows the control and treatment population to be more exchangeable. A wider bandwidth reduces variance but reduces the chance of the estimated effect to be interpreted as the causal effect. Our leave-one-out cross validation procedure suggests that the 1 hour bandwidth (two hours in total) offers a good fit to data. We also consider that the chosen bandwidth is a good option by inspecting the figures for average outcomes at the cutoff and given the limited number of ED encounters around the cutoff.

In this section, we explore the sensitivity of RD estimates to changes in the bandwidth size. Table~\ref{tab:multiple_bandwidth} contains the RD estimates for bandwidth $h$ varied from 0.5 hours to 3 hours. Notoriously, the average time to be roomed is only significant for bandwidth less than 2 hours. On the other hand, ED time to disposition estimates are stable across all bandwidths. Our results support the hypothesis that split flow model reduces ED time to dispositions.  The estimates vary from 20.4 minutes (95\% CI:[-34.7, -6.2]) for the smallest bandwidth (0.5 hours) to 13.6 minutes (95\% CI:[-19.8, -7.5]). In all scenarios, split flow improves the average ED length of stay of patients.

\begin{table}[t!]
\footnotesize
\centering
\begin{tabular}{l c c c c c c } 
\toprule
& \multicolumn{6}{c}{Bandwidth (hours)} \\
\cmidrule(lr){2-7}
Outcome & 0.5 & 1 & 1.5 & 2 & 2.5 & 3 \\
\midrule
& \multicolumn{6}{c}{Including covariates}\\
\cmidrule(lr){2-7}
\multirow{2}{*}{\makecell{  \textbf{Time to be roomed (min.)} }} & 9.1 & 4.6 & 2.2 & 0.7 & -0.01 & -0.5 \\
&  (6.7, 11.4) &  (2.9, 6.2) &  (0.8, 3.5) & (-0.4, 1.8) & (-1.0, 1.0) &  (-1.3, 0.4)    \\
\multirow{2}{*}{\makecell{  \textbf{Time to disposition (min.)} }} & -20.4 & -14.4 & -14.8 & -13.8 & -13.6 & -13.6 \\
& (-34.7, -6.2) & (-24.7, -4.1) & (-23.4,-6.3) & (-21.2, -6.4) & (-20.3, -6.8) & (-19.8, -7.5)\\
\multirow{2}{*}{\makecell{  \textbf{Admission decision (\%)} }} &  -7.9 & -5.8 & -3.6 & -2.8 & -1.9 & -1.9 \\
&  (-12.8, -3.0) & (-9.4, -2.3) & (-6.6, -0.6) & (-5.4,	-0.2) & (-4.3, 0.3) & (4.1, 0.1)\\
\multirow{2}{*}{\makecell{  \textbf{30 days revisit (\%)} }} &  -1.6 & -0.7 & -1.3 & -1.2 & -0.6 & -1.0 \\
& (-5.4, 2.2) & (-3.5, 2.0) & (-3.6, 0.9) & (-3.3, 0.7) & (-2.5, 1.2) & (-2.7, 0.6)  \\
\multirow{2}{*}{\makecell{  \textbf{log(Time to be roomed (min.))} (\%)}} &  46.4 &  18.7 & 7.2 & -0.2 & -0.9 & -2.2 \\
&  (32.5, 61.7) &  (10.0, 28.1) &  (0.5, 14.3) & (-5.7, 5.6) & (-5.8, 4.3) &  (-6.7, 2.4)    \\
\multirow{2}{*}{\makecell{  \textbf{log(Time to disposition (min.))} }} & -12.1 & -9.0 & -9.0 & -8.9 & -8.8 & -8.7 \\
&  (-17.8, -6.0) &  (-13.3, -4.4) &  (-12.5, -5.2) & (-12.0, -5.6) & (-11.6, -5.9) &  (-11.3, -6.0)  \\
\midrule
& \multicolumn{6}{c}{No covariates}\\
\cmidrule(lr){2-7}
\multirow{2}{*}{\makecell{  \textbf{Time to be roomed (min.)} }} & 9.2 & 4.7 & 2.2 & 0.7 & 0.0 & -0.4 \\
&  (6.9, 11.6) &  (3.0, 6.3) &  (0.9, 3.6) & (-0.4,	1.8) & (-1.0, 1.0) &  (-1.3, 0.4)   \\
\multirow{2}{*}{\makecell{  \textbf{Time to disposition (min.)} }} & -21.9 & -15.0 & -16.5 & -14.6 & -13.9 & -14.2 \\
& (-36.4, -7.3) & (-25.6, -4.4) & (-25.2, -7.7) & (-22.2, -7.0) & (-20.7, -7.0) & (-20.5, -7.9)\\
\multirow{2}{*}{\makecell{  \textbf{Admission decision (\%)} }} & -9.4 & -6.8 & -4.5 & -3.4 & -2.4 & -2.6 \\
&  (-14.4, -4.4) & (-10.5, -3.1) & (-7.6, -1.4) & (-6.1, -0.7) & (-4.8, 0.0) & (-4.8, -0.3)\\
\multirow{2}{*}{\makecell{  \textbf{30 days revisit (\%)} }} & -1.5 & -0.6 & -1.3 & -1.2 & -0.6 & -1.0 \\
& (-5.3, 2.2) & (-3.4, 2.0) & (-3.6, 0.9) & (-3.3, 0.8) & (-2.5, 1.2) & (-2.7, 0.7)  \\
\multirow{2}{*}{\makecell{  \textbf{log(Time to be roomed (min.))} (\%)}} &  47.5 &  19.2 & 7.4 & -0.1 & -0.8 & -2.2 \\
&  (33.5, 62.9) & (10.5, 28.6) & (0.7, 14.5) & (-5.6, 5.7) & (-5.8, 4.3) &  (-6.7, 2.5)    \\
\multirow{2}{*}{\makecell{  \textbf{log(Time to disposition (min.))} }} & -12.9 & -9.4 & -9.8 & -9.3 & -8.9 & -9.0 \\
&  (-18.7, -6.6) &  (-13.8, -4.7) & (-13.4, -6.0) & (-12.5, -6.0) & (-11.8, -6.0) &  (-11.7, -6.2)  \\
\bottomrule
\end{tabular}
\caption{Sensitivity to bandwidth size and covariates. RD average treatment effect for multiple outcomes and bandwidths varied from 0.5 hours to 3 hours.}\label{tab:multiple_bandwidth} 
\end{table}

\newpage

\section{RD Figures}\label{app:rd_figures}

\begin{figure}[ht!]
\centering
\includegraphics[width=\textwidth]{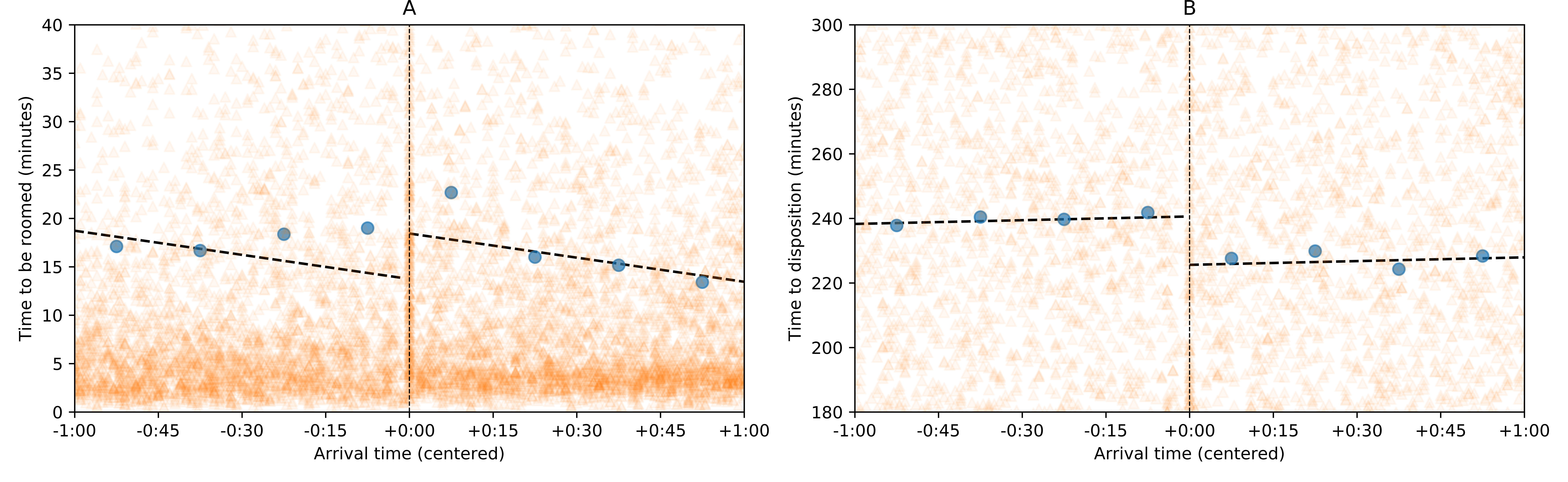}
\caption[]{RD for continuous outcomes. Average time to be roomed (blue dots, Panel A)  and average time to disposition (Panel B) for different visits (orange triangles) grouped in 15 minutes bins and fitted regression lines (dashed line). Arrival hour is centered at the start of split flow.  Causal effects are measured as the discontinuity between estimated left and right regression models at the split flow start time.}
\label{fig:rd_time_outcomes}

\end{figure}

\begin{figure}[ht!]
\centering
\includegraphics[width=\textwidth]{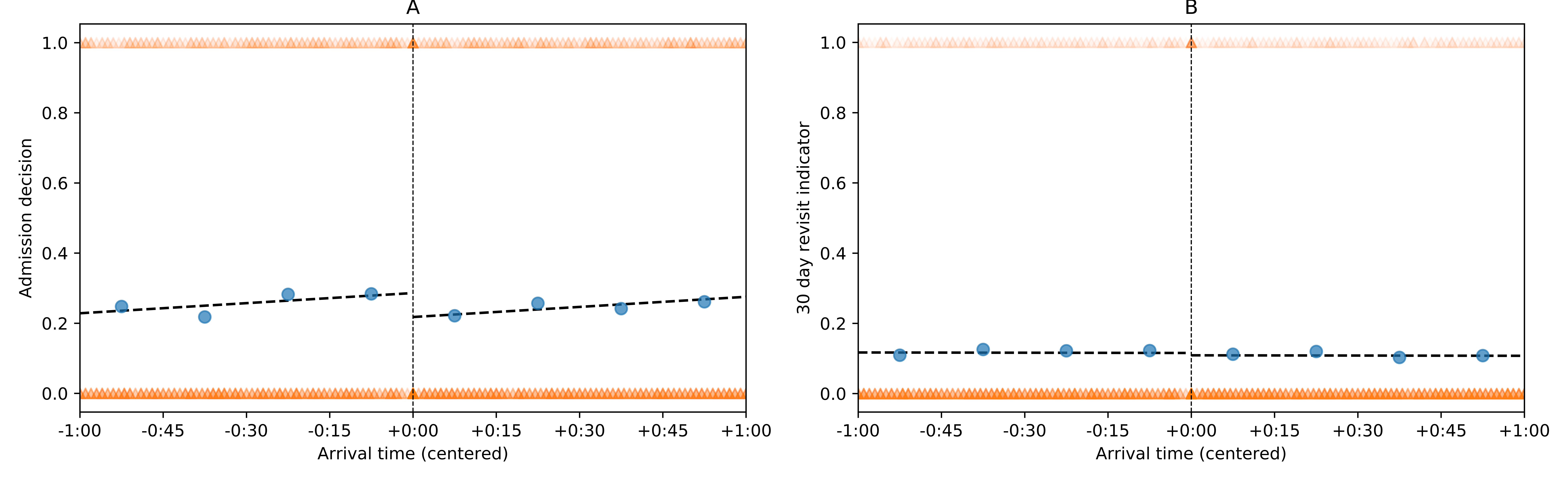}
\caption[]{RD for binary outcomes. Average number of admissions (blue dots, Panel A) and average number of revisits (Panel B) for different ED encounters (orange triangles) grouped in 15 minutes bins and fitted regression lines (dashed line). Arrival hour is centered at the start of split flow.  Causal effects are measured as the discontinuity between estimated left and right regression models at the split flow start time.}
\label{fig:rd_binary_outcomes}

\end{figure}

In this section, we illustrate the RD estimates found in Section~\ref{sec:continuos_results} and Section~\ref{sec:binary_results}. Each figure depicts the average value of outcomes (in blue) for multiple ED encounters (orange triangles) inside the 1 hour bandwidth used for the RD analysis. Fitted regression lines (dashed lines) are also shown on the left and the right of start hour of the split flow (vertical dashed line). Causal effects are measured as the discontinuity formed by left and right regressions at the cutoff. Figure~\ref{fig:rd_time_outcomes} illustrate the results for continuous outcomes, namely, time to be roomed (Panel A) and time to disposition (Panel B). We observe a negative impact of the split flow model on average time to be roomed and a positive effect on average time to disposition at the cutoff. Figure~\ref{fig:rd_binary_outcomes} depicts the analogue for binary outcomes, i.e. admission decision (Panel A) and 30 day revisits (Panel B). We note a significant discontinuity for the number of admissions and a no significant jump for revisits.

\newpage

\section{Moderation analysis for binary outcomes}\label{app:moderation_binary}
We next investigated the influence of moderator variables (i.e., congestion, day of the week, physician workload, and start time of the split flow model) on the effect of split flow on admission decision rates and revisits within 30 days of discharge rates. When adding, one at a time, each possible moderator and their interaction with the split flow indicator to the baseline regression model and testing whether simultaneously the interactions terms are significant, we found that day of the week, physician workload and split flow start hours did not yield a significant results in the model for either admission decision or 30 days revisit rates  (Table~\ref{tab:moderation_binary_one_by_one}). Therefore, day of the week, physician workload and split flow start hours do not appear to moderate the effect of split flow on downstream outcomes. Similarly, congestion was not significant in the model for 30 days revisits (\textit{P} = 0.91) indicating that congestion is not a variable moderating the effect of split flow on 30 days revisit rates.

\begin{figure}[ht!]
\centering
\includegraphics[width=\textwidth]{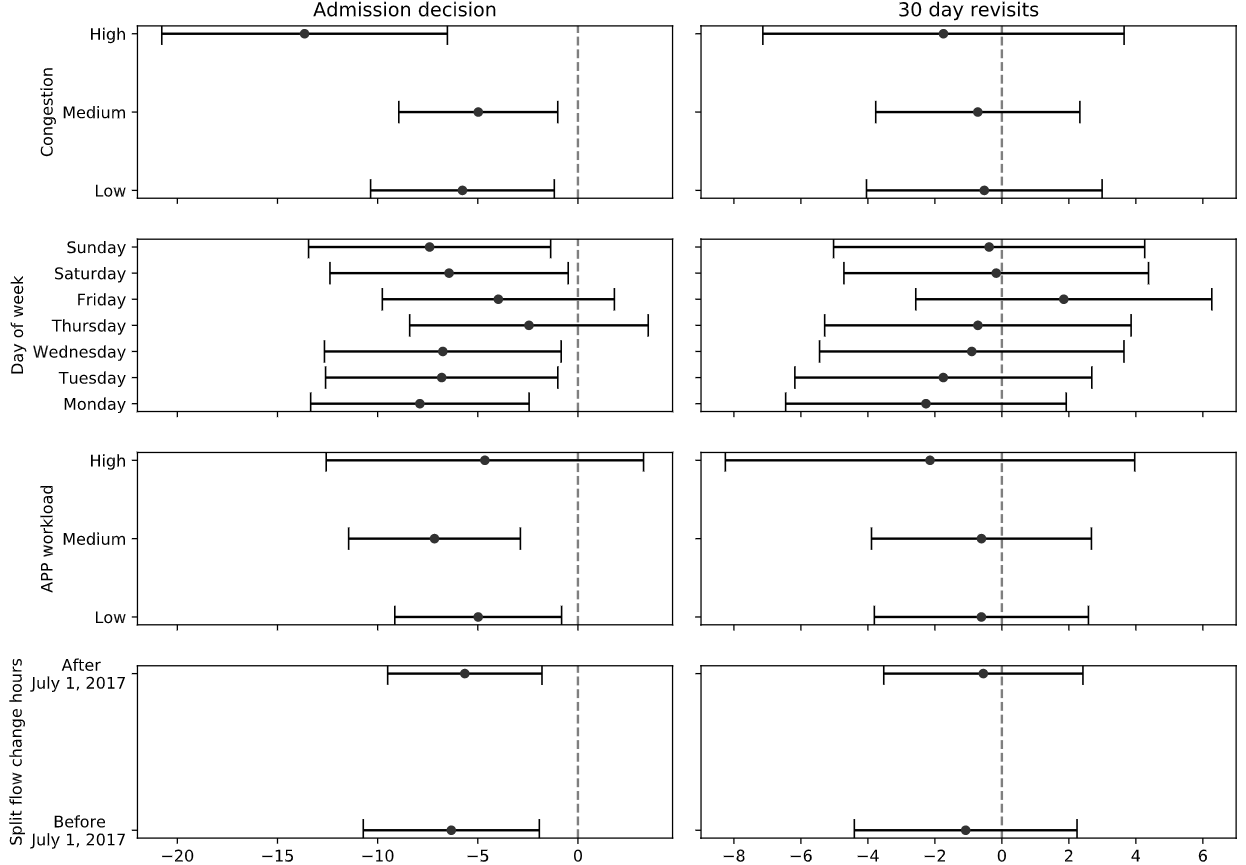}
\caption[]{Average effect (\%) of split flow on admission decision (left) and 30 days revisits (right) during each level of moderator variables: congestion, day of week, physician workload and split flow start hours.}
\label{fig:average_effect_binary_outcomes}
\end{figure}

In contrast, the only moderator that yield significant results was congestion in the model for admission decision (\textit{P} = 0.05). Thus, the effectiveness of the split flow model does appear to be moderated by congestion. Upon closer examination, we find that the split flow model decreases average admission rates during the highest congestion level, and to a lesser degree, during the medium and low congestion levels (Figure~\ref{fig:average_effect_binary_outcomes}).All this might suggest that fewer patients are being admitted under the split flow, especially under high congestion, without impacting revisit rates. 

As we did with continuous outcomes, we also estimate a model including all moderators and their pairwise interactions with the split flow indicators. In this scenario, we still find that day of week, physician workload, and split flow start hours did not yield significant interaction terms (\textit{P} < 0.05 for all outcome, see Table~\ref{tab:moderation_binary_full_model}) for either admission decision and 30 days revisit rates. Surprisingly, congestion was not significant in the full model for either admission decision and 30 days revisits. These findings reinforce the observation that, except for possibly congestion, downstream consequences might be not affected by the operational context of the ED.

\begin{table}[ht!]
    \centering
    \begin{tabular}{l c c}
    \toprule
            \textbf{Moderator} & \textbf{Admission decision} & \textbf{30-days revisit} \\
        \midrule
        Congestion & 0.05 & 0.91\\
        Day of week & 0.72 & 0.81\\
        Physician workload & 0.66 & 0.61\\
        Split flow change hours & 0.74 & 0.73\\
    \bottomrule
    \end{tabular}
    \caption{$P$ values for the joint test of interaction terms of moderator variables with split flow equal to zero for the regression model where each moderator is added one by one.}
    \label{tab:moderation_binary_one_by_one}
\end{table}

\begin{table}[ht!]
    \centering
    \begin{tabular}{l l c c c c}
        \toprule
        & & \multicolumn{2}{c}{\textbf{Admission decision}} & \multicolumn{2}{c}{\textbf{30 days revisit}}\\
        \cmidrule(lr){3-4} \cmidrule(lr){5-6}
         \textbf{Moderator} &  Level & Interaction (95\% CI) & $P$ & Interaction (95\% CI) & $P$\\
        \midrule
         Congestion & Medium & -0.1 (-5.4	5.2) &  \multirow{2}{*}{\makecell{ 0.07 }} & 0.6 (-3.5, 4.6) & \multirow{2}{*}{\makecell{ 0.90 }}\\
         & High & -8.9 (-17.8, -0.1) & & -0.6 (-5.6, 4.5)\\
         \cmidrule(lr){2-6}
         Day of week & Monday & 0.0 (-7.1, 7.1) & \multirow{6}{*}{\makecell{ 0.67 }} & -1.7 (-7.2, 3.7) & \multirow{6}{*}{\makecell{ 0.84 }}\\
         & Tuesday & 0.9 (-6.4, 8.2) & & -1.6 (-7.2, 4.0)\\
         & Wednesday & 1.1 (-6.3, 8.4) & & -0.6 (-6.2, 5.0)\\
         & Thursday & 5.6 (-1.8, 13.1) & & -0.5 (-6.2, 5.2)\\
         & Friday & 3.8 (-3.4, 11.0) & & 2.1 (-3.4, 7.6)\\
         & Saturday & 0.7 (-6.5, 7.9) & & 0.1 (-3.4, 7.6)\\
         \cmidrule(lr){2-6}
         Physician workload & Medium & 1.6 (-3.4, 6.6) & \multirow{2}{*}{\makecell{ 0.81 }} & -1.1 (-4.9, 2.8) & \multirow{2}{*}{\makecell{ 0.86 }}\\
         & High & 1.5 (-3.4, 6.6) & & -0.6  (-5.6, 4.5) \\
         \cmidrule(lr){2-6}
         Split flow start hour & After July 1, 2017 & 0.1 (-4.0, 4.2) & 0.96 & 0.2 (-2.9, 3.3) & 0.90\\
         \bottomrule
    \end{tabular}
    \caption{Estimates (in \%) for the coefficients of the interaction terms between split flow and moderator variables for the regression model including all moderators, and $P$ values for the null hypothesis testing whether the interactions of a given moderator are zero simultaneously.}
    \label{tab:moderation_binary_full_model}
\end{table}

\newpage
\section{Moderation analysis for congestion in relative scale}
\begin{figure}[ht!]
\centering
\includegraphics[width=\textwidth]{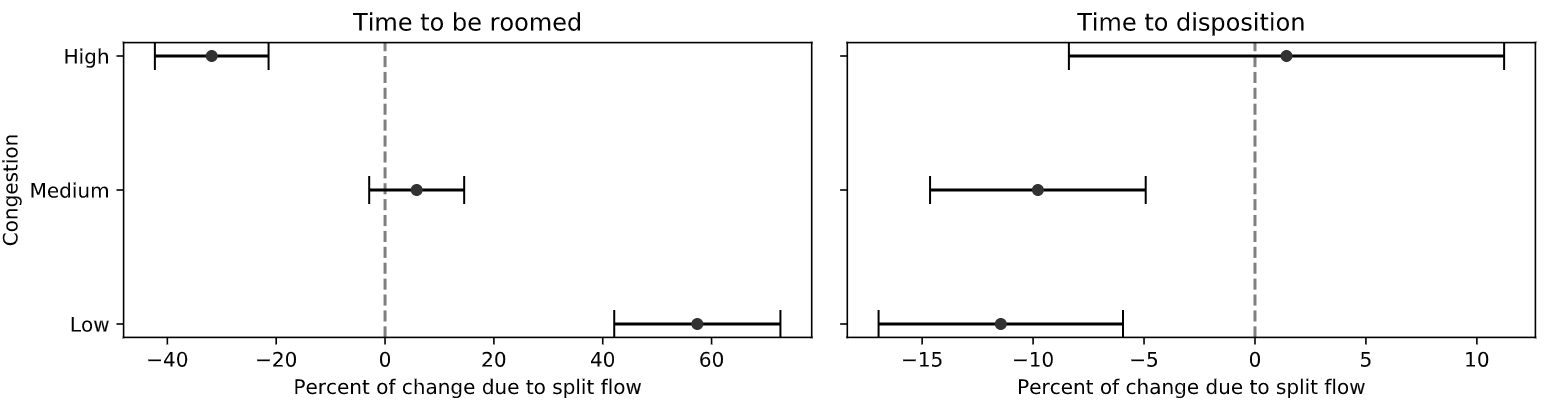}
\caption[]{Average percent of change due to split flow on time to be roomed) and time to disposition during each congestion level.}
\label{fig:moderation_relative_scale}
\end{figure}

In this section we examine the moderation effects of congestion on log transformed time to be roomed and time to disposition. As is standard practice, we take natural logarithm in order to account for the heavy tail of the distribution of length of stay. We find that split flow reduces the average time to be roomed by 31.8\% (95\% CI:[20.6\%, 41.4\%]) during the highest congestion level compared with the traditional nurse-led triage (Table \ref{tab:numbers_moderation_relative_scale}). Split flow, however, increases by a greater percent of 42.8\% (95\% CI:[42.8, 73.3]) the average time to be roomed for low congestion levels compared with the traditional flow. As for time to disposition, we find that split flow reduces average time to disposition for low and middle tertiles by a percent of 11.5\% (95\% CI:[5.7\%, 16.7\%]) and 9.7\% (95\% CI:[4.7\%, 14.5\%]) respectively. These results suggest that the impact of split flow is most notable on time to be roomed compared with time to disposition, in particular, during low congestion levels. This might suggest that split flow actually increases ED length of stay during low congestion levels but this is not the case when we observe the impact in absolute terms (Figure~\ref{fig:moderation_continuous_outcomes}). Similarly, when taken together, we observe percentage reductions of ED length of stay during medium and high levels of congestion. These estimates are depicted in Figure~\ref{fig:moderation_relative_scale}.

\begin{table}[ht!]
    \centering
    \begin{tabular}{l c c}
    \toprule
            \textbf{Congestion level} & \textbf{Time to be roomed} & \textbf{Time to disposition} \\
        \midrule
        High &  -31.8 (-41.4, -20.6) & 1.4 (-7.9, 11.7)\\
        Medium & 5.8 (-2.5, 14.9) & -9.7 (-14.5, -4.7)\\
        Low &  57.3 (42.8, 73.3) & -11.5 (-16.7, -5.7)\\
    \bottomrule
    \end{tabular}
    \caption{Estimates (95\% confidence intervals) for the percent of change on the original scale due to the split flow model on time to be roomed and time to disposition decision after being roomed.}
    \label{tab:numbers_moderation_relative_scale}
\end{table}

\newpage

\section{RD for split flow end hours}\label{app:rd_end_hours}
We have analyzed the effect of the split flow model on outcomes by performing an RD design at the start of the intervention (i.e.,  at 12p). We can also estimate causal effects of split flow on outcomes when split flow ends, i.e., at 9pm.  Three features make the latter analysis different from the one presented in the main text. First, the operational context (E.g., congestion, staffing) of the ED at night is substantially different from the context around noon.  Second, we expect that patients arriving around the start of split flow are different from those arriving at night, right before and right after the end of the intervention. Third, before July 1, 2017, the end of split flow was 10p, which overlaps with the start of a new shift for physicians. Thus, we expect that the effect of split flow at the start will be significantly different from its effect at the end. In particular, we hypothesize that the effect of split flow will be lower at the end of split flow. Here, we test this hypothesis by evaluating impact of split flow on outcomes using an RD design when the forcing variables is centered at the end of the intervention.

As hypothesized, we find that split flow has no significant impact on time to disposition, admission decisions and 30 days revisits (all confidence intervals contain zero) (Table~\ref{tab:rd_end_hours}). By contrast, the average time to be roomed decreases by 3.7 minutes (95\% CI:[1.1, 6.4]).

\begin{table}[ht!]
    \centering
    \begin{threeparttable}
    \begin{tabular}{l c c}
    \toprule
            \textbf{Outcome} & \textbf{Estimate} & \textbf{95\% confidence interval} \\
        \midrule
      Time to be roomed (mins.)& -3.7 & (-6.4, -1.1)\\
      Time to disposition (mins.) & 2.2 & (-8.7, 13.1)\\
      Admission decision (\%) & -0.6 & (-4.4, 3.1)\\
      30 days revisits (\%) & -0.1 & (-3.5, 3.3) \\
      Log(Time to be roomed) (\%)\tnote{1} & -23.3 & (-30.4, -15.0)\\
      Log(Time to disposition) (\%)\tnote{1} & 0.4 & (-5.2, 6.5)\\
    \bottomrule
    \end{tabular}
    \begin{tablenotes}
     \small
     \item[1] Note: Percent of change on the original outcome due to split flow.\\
  \end{tablenotes}
    \end{threeparttable}
    \caption{RD estimates (95\% confidence intervals) the impact of split flow on outcomes when the arrival time is centered at the end of the intervention. A bandwidth size of 1 hour was used (two hours in total). }
    \label{tab:rd_end_hours}
\end{table}

\end{appendices}

\end{document}